\newcounter{myctr}
\def\myitem{\refstepcounter{myctr}\bibfont\noindent\ifnum\themyctr>9\else\phantom{0}\fi\hangindent17pt\themyctr.\enskip}

\documentclass{ws-ijqi}
\usepackage{bm}
\usepackage{graphicx} 
\usepackage{array} 
\usepackage{verbatim} 
\usepackage{subfigure} 
\usepackage{tikz} 
\newcommand{\ket}[1]{\ensuremath{\left|#1\right\rangle}} 
\usepackage{multirow}
\usepackage{marvosym}
\newcommand{\envelope}{(\raisebox{-.5pt}{\scalebox{1.45}{\Letter}}\kern-1.7pt)}
\usepackage{afterpage}
\usepackage{rotating}
	\usepackage{threeparttablex}
	\usepackage{amsmath,amssymb}
	\usepackage{afterpage}
	\usepackage{xcolor}

		\newcommand{\bra}[1]{\ensuremath{\left\langle#1\right|}} 

	\DeclareGraphicsExtensions{.eps}
	\graphicspath{{./figures/}}
	\usepackage{circuitikz}
	\usepackage{xcolor}

\begin{document}
\markboth{Anmer~Daskin, Ananth~Grama, Sabre~Kais}
{Quantum Random State Generation with Predefined Entanglement Constraint}

\catchline{}{}{}{}{}

\title{Quantum Random State Generation with Predefined Entanglement Constraint}

\author{ANMER~DASKIN\footnote{adaskin@purdue.com}\ \and ANANTH~GRAMA}
	\address{Department of Computer Science, Purdue University, West Lafayette, IN, 47907 USA
}

\author{SABRE~KAIS}\address{
Department of Chemistry, Department of Physics and Birck Nanotechnology Center,Purdue University,
West Lafayette, IN 47907 USA;
Qatar Environment and Energy Research Institute, Doha, Qatar
}
	\maketitle
\begin{history}
\received{Day Month Year}
\revised{Day Month Year}
\end{history}
	
	\begin{abstract}
        Entanglement plays an important role in quantum communication, 
	algorithms, and error correction. Schmidt coefficients are correlated to the 
	eigenvalues of the reduced density matrix. These eigenvalues are used in von Neumann
        entropy to quantify the amount of the bipartite entanglement. In this paper,
        we map the Schmidt basis and the associated coefficients to quantum circuits
        to generate random quantum states. We also show that it is possible to adjust
        the entanglement between subsystems by 
	changing the quantum gates corresponding to the Schmidt coefficients. In this manner,
	random quantum states with predefined bipartite entanglement amounts can be generated 
	using random Schmidt basis. This provides a technique for generating equivalent quantum 
	states for given weighted graph states, which are very useful in the study of 
	entanglement, quantum computing, and quantum error correction.
\end{abstract}

\section{Introduction}

In quantum information, a quantum state encodes information and is used in the design of
algorithms. Random numbers and random matrix theory \cite{Edelman2005} play important roles in 
various applications, ranging from wireless communications \cite{Tulino2004random} to
determining physical properties of a quantum system \cite{Beenakker1997}. Consequently,
generating random quantum states is important in quantum communication and information.
For instance, unique random states can be used to design quantum bills (money) \cite{Wiesner1983}. 
Statistical properties of random quantum states show that random quantum states generated 
within some restricted set of states can still be effectively random 
\cite{Wootters1990random}.

A quantum state, defined as a vector in Hilbert space, contains all the accessible-measurable 
information about the system \cite{Sakurai1985modern}. Entanglement is one of the
quantum mechanical accessible phenomena used to build efficient quantum algorithms.
For a given multi-qubit state, a graph state \cite{Hans2001,Dur2003,Nest2004} can be used to specify
the graph-based representation of the entanglement between qubits. A graph state is comprised
of vertices and edges, where the vertices correspond to qubits and edges represent
entanglement between qubits. Non-local properties \cite{Otfried2005,Scarani2005} and 
entanglement characterizations \cite{Hein2004} of graph states have been studied, and their
use has been demonstrated for different applications in quantum error
correction \cite{Schlingemann2001}, quantum communication, and one-way quantum
computation\cite{Raussendorf2001}, among others (Hein et al.\cite{HeinReview} present an
excellent review of the applications of graph states). Realization of 
graph states has been experimentally demonstrated for six photons \cite{lu2007experimental}.
As a generalization of graph states, weighted graph states include weights on each edge,
quantifying the amount of the entanglement. Weighted graph states are shown to be useful in 
the study of bipartite entanglement in spin chains \cite{Hartmann2005} and many-body quantum 
states \cite{Hartmann2007}. Based on a weighted graph state representation of certain 
classes of multi-particle entangled states, a variational method \cite{Anders2006,Anders2007} is 
proposed for arbitrary spin and infinite-dimensional systems. These representations are
also used in error correction schemes in one-way quantum computing \cite{Campbell2007},
and in many other applications (please refer to Hein at al.\cite{HeinReview}).
 
In this paper, we map the Schmidt basis and the associated coefficients to quantum circuits
to generate random quantum states. We show that for state generation, by using 
quantum gates corresponding to Schmidt coefficients, the amount of the bipartite 
entanglement between subsystems can be controlled. Therefore, we show that if quantum 
gates corresponding to the Schmidt basis are chosen randomly, one can generate random states 
with bipartite entanglement amounts predefined by the gates implementing the coefficients. 
This provides a way to tune the entanglement between subsystems in a generated state, which 
can be used to generate certain type of weighted graph states on quantum computers. This
can be used in the utilization and characterization of entanglement \cite{White1999} in 
quantum communication, cryptography, and cluster state computation \cite{Briegel2001}.
In addition, our method can be used to simulate entanglement distribution of particular
quantum systems in quantum computing. An example of this is in the simulation of the entanglement
distribution in light-harvesting complexes\cite{Whaley2010,Fassioli2010} to investigate
energy transfer and efficiency.

\section{Preliminaries}

\paragraph{Schmidt Decomposition}
Given Hilbert spaces $H_A$ and $H_B$ of dimension $d_A$ and $d_B$,
and a quantum state $\ket{\psi}\in H_{AB}=H_A\otimes H_B$,
the Schmidt decomposition is defined as:
\begin{equation}
\ket{\psi}=\sum_i^{min(d_A,d_B)}s_i\ket{u_i}\ket{v_i},
\end{equation}
where $s_i$s are the Schmidt coefficients, and $\ket{u_i}$ and $\ket{v_i}$ are the state vectors that form the Schmidt bases in $H_A$ and $H_B$, respectively.
The reduced density matrix for system $A$ or $B$ can be found from the Schmidt
decomposition as follows:
\begin{equation}
\rho_A = \sum_i s_i^2 \ket{u_i}\bra{u_i}.
\end{equation}
The above expression shows that the coefficients of the Schmidt decomposition
are related to the eigenvalues of the reduced density matrix. 

\paragraph{Von Neumann Entropy}
For a given density matrix $\rho$, the von Neumann Entropy is defined as:
\begin{equation}
S(\rho)=-Tr(\rho\ln{\rho}),
\end{equation}
where the notation $Tr$ describes the trace of a matrix.
If the density matrix $\rho$ with eigenvalues $\lambda_j$ and associated eigenvectors $\ket{j}$ has the eigenvalue decomposition
$\rho=\sum_j\lambda_j\ket{j}\bra{j}$, then the entropy can be defined as:
\begin{equation}
S(\rho)=-\sum_j\lambda_j\ln{\lambda_j}
\end{equation}
For pure states, we can use the Schmidt coefficients in the von Neumann Entropy
to quantify the bipartite entanglement between systems $A$ and $B$ as:
\begin{equation}
S(\rho_A)=S(\rho_B)=-\sum_j^{min(d_A,d_B)}s_j^2\ln{s_j^2},
\end{equation}
where $\rho_A$ and $\rho_B$ are the density matrices for the systems $A$ and $B$, respectively.

\section{Random State Generation With Predetermined Entanglement}

Since the Schmidt coefficients are important in determining entanglement, by suitably mapping
the Schmidt decomposition to a circuit design, we can control entanglement. Please note
that in this paper, for simplicity,  we will only consider the real space for the circuit designs, but they can be generalized to complex space.

\subsection{2-qubit Case}
  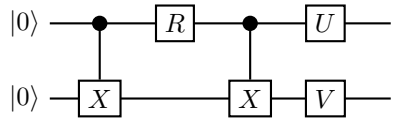
\begin{figure}[ht]
\centerline{
    \begin{tikzpicture}[thick,stepy=3mm]
    %
    \tikzstyle{operator} = [draw,fill=white,minimum size=1.0em] 
    \tikzstyle{control} = [fill,draw=black!105, shape=circle,minimum size=5pt,inner sep=0pt]	
\tikzstyle{operator2} = [fill=white,draw, shape=circle,minimum size=1em,inner sep=0pt]	
  \tikzstyle{cross} = [fill,draw=black!105, shape=cross out,rotate=45,minimum size=5pt,inner sep=0pt]
\tikzstyle{zerocontrol}= [fill=white,draw=black!105,shape=circle,minimum size=5pt,inner sep=0pt]	
    \tikzstyle{surround} = [fill=blue!10,thick,draw=black,rounded corners=1mm]
    %
    \node at (0,0) (q1) {\ket{0}};
    \node at (0,-1) (q2) {\ket{0}};
\node[] (edge1) at (5,-0){} edge[-](q1);
\node[] (edge2) at (5,-1){} edge[-](q2);
    \node[operator] (op41) at (4,-1) {$V$};
   \node[operator] (op41) at (4,-0) {$U$};
\node[operator](op43) at(3,-1){$X$};
\node[control](op42) at (3,-0){} edge[-](op43);
    \node[operator] (op42) at (2,-0) {$R$};
\node[operator](op43) at(1,-1){$X$};
\node[control](op42) at (1,-0){} edge[-](op43);
    \end{tikzpicture}
  }
\caption{Quantum circuit which is found by following the Schmidt decomposition and can generate any quantum state of dimension 4. In the circuit $X$ represents a quantum $NOT$ gate, $U$ and $V$ are the Schmidt basis and R is the rotation gate defined with the Schmidt coefficients. }
\label{fig:schmidt2}
\end{figure}
The Schmidt decomposition for $H=H_1\otimes H_2$, and
$H_1$ and $H_2\in \mathcal{R}^{\otimes^1}$ is as follows:
\begin{equation}
\ket{\psi}=\sum_{i=1}^{2}s_i\ket{u_i}\ket{v_i}.
\end{equation} 
The circuit in Fig.\ref{fig:schmidt2} can be used generate any general $\ket{\psi}$
state for two qubits, where the entanglement defined by  the quantum gate $R$
whose elements are determined from the Schmidt coefficients. In the figure $X$ is the quantum $NOT$ gate, $U$ and $V$ are the Schmidt basis, and the elements of $R$ are  the Schmidt coefficients determining the entanglement: 
\begin{equation}
R=\left(\begin{matrix}
s_1 &-s_2\\s_2 &s_2
\end{matrix} \right).
\end{equation}
Choosing the elements $s_1$ and $s_2$, which are the Schmidt coefficients, and random Schmidt basis $U$ and $V$, 
one can also create a two-qubit random state with predetermined entanglement. 

\subsection{Generalization to $n$ qubits}

We can generalize the idea to an $n$-qubit system:
The Kronecker tensor product of the Schmidt bases $U$ and $V$ can be written in matrix form as: 
\begin{equation}
U \otimes V =
\left[\begin{matrix}
u_{\bullet1}\otimes v_{\bullet1}& \dots &u_{\bullet1}\otimes v_{\bullet k}&
 u_{\bullet2}\otimes v_{\bullet1}& \dots& u_{\bullet2}\otimes v_{\bullet k}& \dots &
 u_{\bullet k}\otimes v_{\bullet1}& \dots& u_{\bullet k}\otimes v_{\bullet k} 
\end{matrix}\right],
\end{equation}
where $u_{\bullet i}$ and $v_{\bullet j}$ represent the $i$th and $j$th column of $U$ and $V$, respectively, and $k$ represents the number of columns.
In the Schmidt decomposition of a vector $\ket\psi$:
\begin{equation}
\ket\psi= \sum_{i=1}^k s_i u_{\bullet i} \otimes v_{\bullet i},
\end{equation}
the Schmidt coefficients $s_1 \dots s_k$ are related to the columns:
$1, (k+2),(2k+3), \dots, (k^2)$, respectively.
Therefore, if we have an input state $\ket\varphi$ to  $(U\otimes V)$ in
the following form:
\begin{equation}
\label{Eq:varphi}
\ket\varphi =\left[
\begin{smallmatrix}
 s_1\\
 0\\
 \vdots\\
 0\\
 s_2\\
 0\\
 \vdots\\
 0\\
 s_3\\
 0\\
 \vdots\\
 0\\
 s_k
\end{smallmatrix}\right],
\end{equation}
then $(U\otimes V) \ket\varphi =\ket\psi=  \sum_{i=1}^ks_i u_{\bullet i} \otimes v_{\bullet i}$.
If we assume the initial input to the circuit is  $\ket{\bm{0}}$, then the first column of
the matrix representation of the circuit defines the output.
Therefore, to generate $\ket\varphi$, first we construct the Schmidt coefficients in
the first column of the matrix $S$ of dimension $min(d_A, d_B)$. If $S$ is on the
first subsystem, then the global unitary operator is  $(S\otimes I)$ with the first column:
\begin{equation}
\left[
\begin{smallmatrix}
 s_1\\
 0\\
 \vdots\\
 0\\
 s_2\\
 0\\
 \vdots\\
 0\\
 s_3\\
 0\\
 \vdots\\
 0\\
 s_k
 \\0\\ \vdots\\ 0
\end{smallmatrix}\right]
\end{equation}
To get the Schmidt coefficients to the rows $1, (k+2),(2k+3), \dots, (k^2)$ as
in Eq.(\ref{Eq:varphi}), we apply a permutation matrix $P$ to switch the rows and
columns: $P(S\otimes I)P$. Therefore, the final circuit can be represented by
the matrix vector product as:
 \begin{equation}
 \ket\psi=(U\otimes V )P(S\otimes I)P\ket{\bm{0}}.
 \end{equation}
In the corresponding circuit design, $U$ and $V$ are defined as the operators
on the first and the second subsystems, respectively. $S$, whose first column is
the Schmidt coefficients, is considered on the first subsystem. Since the operator
$P$ is a permutation matrix, it can be implemented by a combination of controlled $NOT$ ($CNOT$) gates.

\subsection{4-qubit Case}
As an example, consider a 4-qubit system, where the subsystems 
are composed of two qubits: $H_{12}$ for the first and second qubits and $H_{34}$ for the third and fourth qubits. Thus, in the Schmidt decomposition, there are four
coefficients: $s_1,s_2,s_3,$ and $s_4$. The circuit in Fig.\ref{fig:schmidt4}
generates any quantum state that has the Schmidt coefficients implemented by $S$
and Schmidt bases implemented by $U$ and $V$. For the implementation of $S$
given in Fig.\ref{fig:s}, we follow the idea first presented in ref.\cite{Daskin2012}:
The coefficients are divided into two unit vectors as
$1/k_1\left[\begin{smallmatrix} s_1 \\ s2\end{smallmatrix}\right]$ and
$1/k_2\left[\begin{smallmatrix} s_3 \\ s4\end{smallmatrix}\right]$, with normalization
constants $1/k_1$ and $1/k_2$. Then, the rotation gates in Fig.\ref{fig:s} are defined as:
\begin{equation}
R_1=\frac{1}{k_1}\left(\begin{matrix}
s_1 &-s_2\\
s_2& s_1
\end{matrix}\right),\ 
R_2=\frac{1}{k_2}\left(\begin{matrix}
s_3 &-s_4\\
s_4& s_3
\end{matrix}\right),
\end{equation}
and
\begin{equation}
R_3=\left(\begin{matrix}
k_1& -k_2\\
k_2 &k_1
\end{matrix} \right).
\end{equation}
$S$ is constructed using the above rotation gates as:
\begin{equation}
S=\left(\begin{matrix}R_1& I\\ I&R_2\end{matrix}\right)(R_3 \otimes I).
\end{equation}
The first column of $S$ consists of the Schmidt coefficients, which is shown in Fig.\ref{fig:s}.

  \begin{figure*}[Ht]
\centerline{
    \begin{tikzpicture}[thick,stepy=3mm]
    %
    \tikzstyle{operator} = [draw,fill=white,minimum size=1.0em] 
    \tikzstyle{control} = [fill,draw=black!105, shape=circle,minimum size=5pt,inner sep=0pt]	
\tikzstyle{operator2} = [fill=white,draw, shape=circle,minimum size=1em,inner sep=0pt]	
\tikzstyle{operatorbig} = [draw,fill=white,minimum size=4.0em] 
  \tikzstyle{cross} = [fill,draw=black!105, shape=cross out,rotate=45,minimum size=5pt,inner sep=0pt]
\tikzstyle{zerocontrol}= [fill=white,draw=black!105,shape=circle,minimum size=5pt,inner sep=0pt]	
    \tikzstyle{surround} = [fill=blue!10,thick,draw=black,rounded corners=1mm]
    %
    \node at (-0,0) (q1) {\ket{0}};
    \node at (-0,-1) (q2) {\ket{0}};
      \node at (-0,-2) (q3) {\ket{0}};
    \node at (-0,-3) (q4) {\ket{0}};
\node (l1) at (12.5,0){};
\node[] (edge1) at ($(q1)+(l1)$){} edge[-](q1);
\node[] (edge2) at ($(q2)+(l1)$){} edge[-](q2);
\node[] (edge2) at ($(q3)+(l1)$){} edge[-](q3);
\node[] (edge2) at ($(q4)+(l1)$){} edge[-](q4);
\node (c) at (11.5,0){};
    \node[operatorbig] (op1) at ($1/2*(q1)+1/2*(q2)+(c)$) {$U$};
    \node[operatorbig] (op2) at ($1/2*(q3)+1/2*(q4)+(c)$) {$V$};
\node (c) at (10,0){};
\node[operator] (op1) at ($(q3)+(c)$) {$X$};
\node[control](op0) at ($(q2)+(c)$){} edge[-](op1);
\node[zerocontrol](op1) at ($(q1)+(c)$){} edge[-](op0);
\node (c) at (9,0){};
\node[operator] (op1) at ($(q4)+(c)$) {$X$};
\node[control](op0) at ($(q2)+(c)$){} edge[-](op1);
\node[zerocontrol](op1) at ($(q1)+(c)$){} edge[-](op0);
\node (c) at (8,0){};
\node[operator] (op1) at ($(q3)+(c)$) {$X$};
\node[control](op0) at ($(q2)+(c)$){} edge[-](op1);
\node[control](op1) at ($(q1)+(c)$){} edge[-](op0);
\node (c) at (7,0){};
\node[operator] (op1) at ($(q4)+(c)$) {$X$};
\node[control](op0) at ($(q2)+(c)$){} edge[-](op1);
\node[control](op1) at ($(q1)+(c)$){} edge[-](op0);
\node (c) at (5.5,0){};
\node[operatorbig] (op1) at ($1/2*(q1)+1/2*(q2)+(c)$) {$S$};
\node (c) at (4,0){};
\node[operator] (op1) at ($(q3)+(c)$) {$X$};
\node[control](op0) at ($(q2)+(c)$){} edge[-](op1);
\node[zerocontrol](op1) at ($(q1)+(c)$){} edge[-](op0);
\node (c) at (3,0){};
\node[operator] (op1) at ($(q4)+(c)$) {$X$};
\node[control](op0) at ($(q2)+(c)$){} edge[-](op1);
\node[zerocontrol](op1) at ($(q1)+(c)$){} edge[-](op0);
\node (c) at (2,0){};
\node[operator] (op1) at ($(q3)+(c)$) {$X$};
\node[control](op0) at ($(q2)+(c)$){} edge[-](op1);
\node[control](op1) at ($(q1)+(c)$){} edge[-](op0);
\node (c) at (1,0){};
\node[operator] (op1) at ($(q4)+(c)$) {$X$};
\node[control](op0) at ($(q2)+(c)$){} edge[-](op1);
\node[control](op1) at ($(q1)+(c)$){} edge[-](op0);
    \end{tikzpicture}
  }
\caption{Quantum circuit for 4 qubits which is found by following the Schmidt decomposition and can generate any quantum state of dimension 16. In the circuit $U$ and $V$ are the Schmidt basis and S implements the Schmidt coefficients controlling the entanglement between $H_{12}$ and $H_{34}$. }
\label{fig:schmidt4}
\end{figure*}
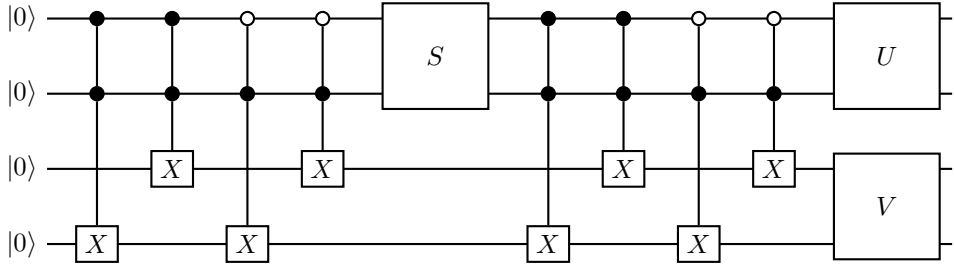
  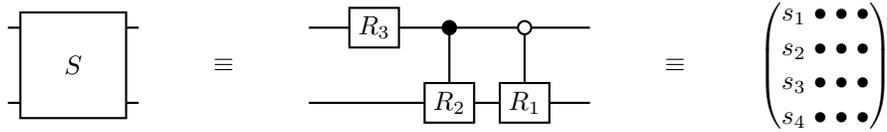
\begin{figure*}
\centerline{
    \begin{tikzpicture}[thick,stepy=3mm]
    %
    \tikzstyle{operator} = [draw,fill=white,minimum size=1.0em] 
    \tikzstyle{control} = [fill,draw=black!105, shape=circle,minimum size=5pt,inner sep=0pt]	
\tikzstyle{operator2} = [fill=white,draw, shape=circle,minimum size=1em,inner sep=0pt]	
\tikzstyle{operatorbig} = [draw,fill=white,minimum size=4.0em] 
  \tikzstyle{cross} = [fill,draw=black!105, shape=cross out,rotate=45,minimum size=5pt,inner sep=0pt]
\tikzstyle{zerocontrol}= [fill=white,draw=black!105,shape=circle,minimum size=5pt,inner sep=0pt]	
    \tikzstyle{surround} = [fill=blue!10,thick,draw=black,rounded corners=1mm]
    %
    \node at (-0,0) (q1) {};
    \node at (-0,-1) (q2) {};
\node (l1) at (2,0){};
\node[] (edge1) at ($(q1)+(l1)$){} edge[-](q1);
\node[] (edge2) at ($(q2)+(l1)$){} edge[-](q2);
\node (c) at (1,0){};
\node[operatorbig] (op1) at ($1/2*(q2)+1/2*(q1)+(c)$) {$S$};
\node (c) at (3,0){};
\node at ($1/2*(q2)+1/2*(q1)+(c)$ (equiv) {$\equiv$};
    \node at (4,0) (q1) {};
    \node at (4,-1) (q2) {};
\node (l1) at (4,0){};
\node[] (edge1) at ($(q1)+(l1)$){} edge[-](q1);
\node[] (edge2) at ($(q2)+(l1)$){} edge[-](q2);
\node (c) at (3,0){};
\node[operator] (op1) at ($(q2)+(c)$) {$R_1$};
\node[zerocontrol](op0) at ($(q1)+(c)$){} edge[-](op1);
\node (c) at (2,0){};
\node[operator] (op1) at ($(q2)+(c)$) {$R_2$};
\node[control](op0) at ($(q1)+(c)$){} edge[-](op1);
\node (c) at (1,0){};
\node[operator] (op1) at ($(q1)+(c)$) {$R_3$};
\node (c) at (5,0){};
\node at ($1/2*(q2)+1/2*(q1)+(c)$ (equiv) {$\equiv$};
\node (c) at (7,0){};
\node at ($1/2*(q2)+1/2*(q1)+(c)$) {$\left(\begin{matrix}
s_1&\bullet & \bullet & \bullet \\
s_2&\bullet & \bullet & \bullet \\
s_3&\bullet & \bullet & \bullet \\
s_4&\bullet & \bullet & \bullet \\
\end{matrix}\right)
$};
    \end{tikzpicture}
  }
\caption{Quantum circuit that has the matrix representation whose first column is the Schmidt coefficients. }
\label{fig:s}
\end{figure*}

\section{Bipartite Entanglement Control for $n$ qubits}

We now show that we can sequentially combine the Schmidt decomposition circuits
for two qubits to control the entanglement between various parts of the system
with the rest of the system in the random state: e.g., for a $5$ qubit system,
controlling entanglement between $H_1$ and $H_{2345}$ and  $H_{12}$ and $H_{345}$. 
 
\subsection{Connecting three qubits linearly}

We start with the initial state $\ket{\psi_0}=\ket{000}$.
If we assume the first two qubits are entangled by the circuit in Fig.\ref{fig:schmidt2},
where the Schmidt basis is chosen to be identity: $V_1=U_1=I$, then we get the following:
\begin{equation}
\ket{\psi_1}= s_1\ket{0}\ket{0}\ket{0} + s_2\ket{1}\ket{1}\ket{0},
\end{equation}
where $s_1$ and $s_2$ are Schmidt coefficients. To also entangle the 2nd and 3rd
qubits, we apply the same Schmidt circuit to these qubits:
First, the $CNOT$ gate is applied:
\begin{equation}
\begin{split}
\ket{\psi_2} = &s_1\ket{0}CNOT(\ket{0}\ket{0}) + s_2\ket{1}CNOT(\ket{1}\ket{0})\\
 =& s_1\ket{0}\ket{0}\ket{0} + s_2\ket{1}\ket{1}\ket{1}.
 \end{split}
\end{equation}
Then, we apply the rotation gate $R$, which has the Schmidt coefficients,
$k_1$ and $k_2$, as elements:
\begin{equation}
R_2=\left(\begin{matrix}
k_1 &-k_2\\k_2 &k_1
\end{matrix} \right)
\end{equation}
This generates the following state:
\begin{equation}
\begin{split}
\ket{\psi_3} = & s_1\ket{0}R_2\ket{0}\ket{0} + s_2\ket{1}R_2\ket{1}\ket{1}\\
= & s_1\ket{0}(k_1\ket{0} + k_2\ket{1})\ket{0} + s_2\ket{1}(-k_2\ket{0} + k_1\ket{1})\ket{1}\\
= &s_1k_1\ket{000}+ s_1k_2\ket{010} -s_2k_2\ket{101} + s_2k_1\ket{111}\\
\end{split}
\end{equation}
After the second $CNOT$, the final state becomes: 
\begin{equation}
\ket{\psi_4} = s_1k_1\ket{000}+ s_1k_2\ket{011} -s_2k_2\ket{101} + s_2k_1\ket{110}\\
\end{equation}
Since   $U_2$ and $V_2$ are the local operators, they do not change the entanglement.
Therefore, the entanglement between $H_{12}$ and $H_{3}$, and the entanglement
between $H_1$ and $H_{23}$ can be found from $\ket{\psi_4}$.

The von Neumann entropy $S(\rho_3)= S(\rho_{12})$ defines the entanglement between the
systems  $H_{12}$ and$H_{3}$. For the entropy, the reduced density matrix $\rho_3$
can be found as follows:
\begin{equation}
\label{Eq:reduced1}
\begin{split}
\rho_3 = &Tr_{H_{12}}(\ket{\psi_4}\bra{\psi_4})\\
= & (s_1k_1)^2\ket{0}\bra{0}+ (s_1k_2)^2\ket{1}\bra{1} +(s_2k_2)^2\ket{1}\bra{1}\\ 
+ &(s_2k_1)^2\ket{0}\bra{0}\\
= & \left((s_1k_1)^2 + (s_2k_1)^2 \right)\ket{0}\bra{0}+ \left( (s_1k_2)^2 + (s_2k_2)^2\right)\ket{1}\bra{1} \\
= &(k_1)^2 \ket{0}\bra{0} + (k_2)^2\ket{1}\bra{1}
\end{split}
\end{equation}
Since $S(\rho_3)=(k_1)^2\ln{(k_1)^2}+(k_2)^2\ln{(k_2)^2}$, which is determined solely
from the Schmidt coefficients, the entanglement is controlled as expected.
 
The entanglement between $H_{12}$ and $H_3$ is also defined as $S(\rho_{1})=S(\rho_{23})$.
Here, the reduced density matrix $\rho_1$ can be obtained as follows:
  \begin{equation}
  \label{Eq:reduced2}
  \begin{split}
\rho_1= &Tr_{H_{23}}(\ket{\psi_4}\bra{\psi_4})\\
= & \left((s_1k_1)^2 + (s_1k_2)^2 \right)\ket{0}\bra{0}+ \left( (s_2k_2)^2 + (s_2k_1)^2\right)\ket{1}\bra{1} \\
= &(s_1)^2 \ket{0}\bra{0} + (s_2)^2\ket{1}\bra{1}
\end{split}
  \end{equation}
Finally, we find $S(\rho_1)=(s_1)^2\ln{(s_1)^2}+(s_2)^2\ln{(s_2)^2}$.
Eq.(\ref{Eq:reduced1}) and Eq.(\ref{Eq:reduced2}) prove that we can use the Schmidt
circuit sequentially to achieve desired entanglement between two disentangled subsystems. 
 
\subsection{Definition of a Graph State}

For a given multi-qubit state, a graph state is an instance of the graph-based representation
of the entanglement between qubits\cite{HeinReview}. They are used in determining
the capacity of quantum channels and quantum error correction. We use weighted graph
states, where vertices represent qubits (a vertex can also be a subsystem), and an
edge between vertices $v_i$ and $v_j$ determines the bipartite entanglement between
subsystems $i$ and $j$. 
 
\begin{figure*}
\centerline{
\resizebox{5in}{!}{
    \begin{tikzpicture}[thick,stepy=3mm]
    %
    \tikzstyle{operator} = [draw,fill=white,minimum size=1.0em] 
    \tikzstyle{control} = [fill,draw=black!105, shape=circle,minimum size=5pt,inner sep=0pt]	
\tikzstyle{operator2} = [fill=white,draw, shape=circle,minimum size=1em,inner sep=0pt]	
\tikzstyle{operatorbig} = [draw,fill=white,minimum size=4.0em] 
  \tikzstyle{cross} = [fill,draw=black!105, shape=cross out,rotate=45,minimum size=5pt,inner sep=0pt]
\tikzstyle{zerocontrol}= [fill=white,draw=black!105,shape=circle,minimum size=5pt,inner sep=0pt]	
    \tikzstyle{surround} = [fill=blue!10,thick,draw=black,rounded corners=1mm]
    %
    \node[control] at (0,-2) (q1) {};
      \node[] at (0.2,-1.8) (q11) {1};
 \node[control] at (0,-1) (q2) {}; 
 \node[control] at (0,-3) (q3) {}; 
 \node[control] at (1,-2) (q4) {}; 
     \node[control] at (-1,-2) (q5) {}; 
      \node[] at (-0.2,-1.2) (q21) {2}; 
 \node[] at (0.2,-3.2) (q31) {3}; 
 \node[] at (0.8,-2.2) (q41) {4}; 
     \node[] at (-0.8,-2.2) (q51) {5}; 
\node[] (edge1) at (q5){} edge[-](q1);
\node[] (edge1) at (q4){} edge[-](q1);
\node[] (edge1) at (q3){} edge[-](q1);
\node[] (edge1) at (q2){} edge[-](q1);
\node (c) at (0.5,0){};
\node at ($(q4)+(c)$ (equiv) {$\equiv$};
 \node at (2,0) (q1) {\ket{0}};
\node at (2,-1) (q2) {\ket{0}};
\node at (2,-2) (q3) {\ket{0}};
\node at (2,-3) (q4) {\ket{0}};
\node at (2,-4) (q5) {\ket{0}};
\node (l1) at (16.5,0){};
\node[] (edge1) at ($(q1)+(l1)$){} edge[-](q1);
\node[] (edge2) at ($(q2)+(l1)$){} edge[-](q2);
\node[] (edge3) at ($(q3)+(l1)$){} edge[-](q3);
\node[] (edge4) at ($(q4)+(l1)$){} edge[-](q4);
\node[] (edge5) at ($(q5)+(l1)$){} edge[-](q5);
\node (c) at (4,0){};
 \node[operator] (op41) at ($(q1)+(c)$) {$U_1$};
 \node[operator] (op41) at ($(q2)+(c)$) {$V_1$};
\node (c) at (3,0){};
\node[operator](op43) at($(q2)+(c)$){$X$};
\node[control](op42) at ($(q1)+(c)$){} edge[-](op43);
\node (c) at (2,0){};
    \node[operator] (op42) at ($(q1)+(c)$) {$R_1$};
\node (c) at (1,0){};
\node[operator](op43) at($(q2)+(c)$){$X$};
\node[control](op42) at ($(q1)+(c)$){} edge[-](op43);
\node (c) at (8,0){};
 \node[operator] (op41) at ($(q1)+(c)$) {$U_2$};
 \node[operator] (op41) at ($(q3)+(c)$) {$V_2$};
\node (c) at (7,0){};
\node[operator](op43) at($(q3)+(c)$){$X$};
\node[control](op42) at ($(q1)+(c)$){} edge[-](op43);
\node (c) at (6,0){};
    \node[operator] (op42) at ($(q1)+(c)$) {$R_2$};
\node (c) at (5,0){};
\node[operator](op43) at($(q3)+(c)$){$X$};
\node[control](op42) at ($(q1)+(c)$){} edge[-](op43);
\node (c) at (12,0){};
 \node[operator] (op41) at ($(q1)+(c)$) {$U_3$};
 \node[operator] (op41) at ($(q4)+(c)$) {$V_3$};
\node (c) at (11,0){};
\node[operator](op43) at($(q4)+(c)$){$X$};
\node[control](op42) at ($(q1)+(c)$){} edge[-](op43);
\node (c) at (10,0){};
    \node[operator] (op42) at ($(q1)+(c)$) {$R_3$};
\node (c) at (9,0){};
\node[operator](op43) at($(q4)+(c)$){$X$};
\node[control](op42) at ($(q1)+(c)$){} edge[-](op43);
\node (c) at (16,0){};
 \node[operator] (op41) at ($(q1)+(c)$) {$U_4$};
 \node[operator] (op41) at ($(q5)+(c)$) {$V_4$};
\node (c) at (15,0){};
\node[operator](op43) at($(q5)+(c)$){$X$};
\node[control](op42) at ($(q1)+(c)$){} edge[-](op43);
\node (c) at (14,0){};
    \node[operator] (op42) at ($(q1)+(c)$) {$R_4$};
\node (c) at (13,0){};
\node[operator](op43) at($(q5)+(c)$){$X$};
\node[control](op42) at ($(q1)+(c)$){} edge[-](op43);
    \end{tikzpicture}
  }}
\caption{Quantum circuit that can generate random state for 5 qubits with the entanglement amount between qubits defined on the star graph. }
\label{fig:stargraph}
\end{figure*}
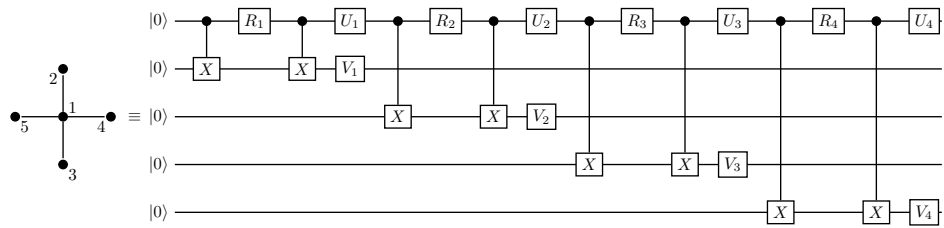

If a graph state is acyclic, i.e. there is only one edge connecting two subsystems,
successively using the Schmidt circuit in Fig.\ref{fig:schmidt2} and controlling the
Schmidt coefficients, as done for three qubits above, the desired entanglement
between each subsystems can be derived. Example circuits are given in
Fig.\ref{fig:lineargraph} and Fig.\ref{fig:stargraph} for linear (path) and star graphs
with five qubits.  A linear graph or path graph consists of vertices and edges that can be drawn as a single straight line where there are two terminal vertices of degree 1 at the beginning and at the end of the line and the remaining vertices are in the middle and have degree 2. A star graph with $n$ vertices have one vertex of degree $n-1$ and all the other vertices have degree 1.  When a star graph is drawn, as its name suggests, it forms a star where the vertex having degree $n-1$ is located in the middle.  Similar circuit designs can be generated for different graphs
in the same manner.

  \begin{figure*}
\centerline{
\resizebox{5in}{!}{
    \begin{tikzpicture}[thick,stepy=3mm]
    %
    \tikzstyle{operator} = [draw,fill=white,minimum size=1.0em] 
    \tikzstyle{control} = [fill,draw=black!105, shape=circle,minimum size=5pt,inner sep=0pt]	
\tikzstyle{operator2} = [fill=white,draw, shape=circle,minimum size=1em,inner sep=0pt]	
\tikzstyle{operatorbig} = [draw,fill=white,minimum size=4.0em] 
  \tikzstyle{cross} = [fill,draw=black!105, shape=cross out,rotate=45,minimum size=5pt,inner sep=0pt]
\tikzstyle{zerocontrol}= [fill=white,draw=black!105,shape=circle,minimum size=5pt,inner sep=0pt]	
    \tikzstyle{surround} = [fill=blue!10,thick,draw=black,rounded corners=1mm]
    %
    \node[control] at (-0,0) (q1) {};
    \node[control] at (-0,-1) (q2) {};
    \node[control] at (-0,-2) (q3) {};
    \node[control] at (-0,-3) (q4) {};
    \node[control] at (-0,-4) (q5) {};    
\node[] (edge1) at (q5){} edge[-](q1);
\node (c) at (1,0){};
\node at ($(q3)+(c)$ (equiv) {$\equiv$};
 \node at (2,0) (q1) {\ket{0}};
\node at (2,-1) (q2) {\ket{0}};
\node at (2,-2) (q3) {\ket{0}};
\node at (2,-3) (q4) {\ket{0}};
\node at (2,-4) (q5) {\ket{0}};
\node (l1) at (16.5,0){};
\node[] (edge1) at ($(q1)+(l1)$){} edge[-](q1);
\node[] (edge2) at ($(q2)+(l1)$){} edge[-](q2);
\node[] (edge3) at ($(q3)+(l1)$){} edge[-](q3);
\node[] (edge4) at ($(q4)+(l1)$){} edge[-](q4);
\node[] (edge5) at ($(q5)+(l1)$){} edge[-](q5);
\node (c) at (4,0){};
 \node[operator] (op41) at ($(q1)+(c)$) {$U_1$};
 \node[operator] (op41) at ($(q2)+(c)$) {$V_1$};
\node (c) at (3,0){};
\node[operator](op43) at($(q2)+(c)$){$X$};
\node[control](op42) at ($(q1)+(c)$){} edge[-](op43);
\node (c) at (2,0){};
    \node[operator] (op42) at ($(q1)+(c)$) {$R_1$};
\node (c) at (1,0){};
\node[operator](op43) at($(q2)+(c)$){$X$};
\node[control](op42) at ($(q1)+(c)$){} edge[-](op43);
\node (c) at (8,0){};
 \node[operator] (op41) at ($(q2)+(c)$) {$U_2$};
 \node[operator] (op41) at ($(q3)+(c)$) {$V_2$};
\node (c) at (7,0){};
\node[operator](op43) at($(q3)+(c)$){$X$};
\node[control](op42) at ($(q2)+(c)$){} edge[-](op43);
\node (c) at (6,0){};
    \node[operator] (op42) at ($(q2)+(c)$) {$R_2$};
\node (c) at (5,0){};
\node[operator](op43) at($(q3)+(c)$){$X$};
\node[control](op42) at ($(q2)+(c)$){} edge[-](op43);
\node (c) at (12,0){};
 \node[operator] (op41) at ($(q3)+(c)$) {$U_3$};
 \node[operator] (op41) at ($(q4)+(c)$) {$V_3$};
\node (c) at (11,0){};
\node[operator](op43) at($(q4)+(c)$){$X$};
\node[control](op42) at ($(q3)+(c)$){} edge[-](op43);
\node (c) at (10,0){};
    \node[operator] (op42) at ($(q3)+(c)$) {$R_3$};
\node (c) at (9,0){};
\node[operator](op43) at($(q4)+(c)$){$X$};
\node[control](op42) at ($(q3)+(c)$){} edge[-](op43);
\node (c) at (16,0){};
 \node[operator] (op41) at ($(q4)+(c)$) {$U_4$};
 \node[operator] (op41) at ($(q5)+(c)$) {$V_4$};
\node (c) at (15,0){};
\node[operator](op43) at($(q5)+(c)$){$X$};
\node[control](op42) at ($(q4)+(c)$){} edge[-](op43);
\node (c) at (14,0){};
    \node[operator] (op42) at ($(q4)+(c)$) {$R_4$};
\node (c) at (13,0){};
\node[operator](op43) at($(q5)+(c)$){$X$};
\node[control](op42) at ($(q4)+(c)$){} edge[-](op43);
    \end{tikzpicture}
  }}
\caption{Quantum circuit that can generate random state for 5 qubits with the entanglement amount between qubits defined on the linear graph. }
\label{fig:lineargraph}
\end{figure*}
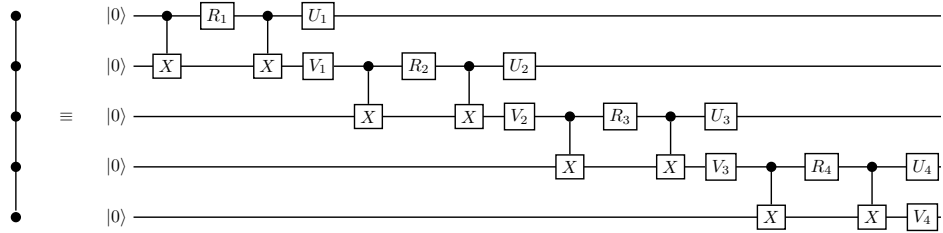

\section{Numerical Results}
\begin{figure}[ht]
\centering
\includegraphics[width=3in]{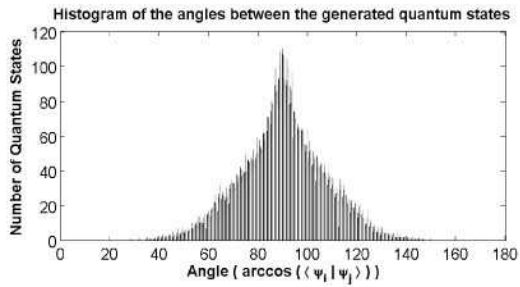}
\caption{Angles between the generated quantum states for an eight-qubit star graph:
Both the Schmidt coefficients and basis are chosen randomly for each state.}
\label{fig:refdistribution}
\end{figure}

The statistical properties of the entanglement of a large bipartite quantum system have been analyzed by Facchi et. al \cite{Facchi2008}. Pasquale et. al \cite{Pasquale2010} have investigated the statistical distribution of the Schmidt coefficients to obtain characterization of the statistical features of the bipartite entanglement of a large
quantum system in a pure state. In addition, the behavior of bipartite entanglement at the fixed von Neumann entropy has been recently studied in ref.\cite{Facchi2013}.
Here, we now show the degree of the randomness of the output states, generated by the circuits described in the previous section, through a given probability distribution of the generated states.  
Let $G(m, n)$ be an $m\times n$ matrix of independent and identically distributed standard
normal real random variables. The distribution of the matrices is defined as \cite{Edelman2005,Stewart1980}:
\begin{equation}
\frac{1}{(2\pi)^{\beta mn/2}}e^{-\frac{1}{2}
||G||_F^2},
\end{equation}
where $||G||_F=\sqrt{Tr(G^*G)}$ is the Frobenius norm of the matrix $G$, and $\beta$ takes values based on considering  real matrices($\beta=1$), the complexes ($\beta=2$), or the quaternions $(\beta=4)$.
In $MATLAB$, we use the function $G=randn(m,n)$ to generate matrices with the above
Gaussian distribution with $\beta=1$. Starting with a normally distributed matrix and taking QR
or singular value decomposition of the matrix generates random orthogonal matrices
distributed according to Haar measure. One can also  generate standard random
orthogonal matrices with the same distribution by using successive plane rotations
with random angle values generated according to Gaussian distribution \cite{Anderson1987}. 
Therefore, for quantum states generated using the Schmidt decomposition by choosing
random Schmidt coefficients and basis (or the corresponding quantum gates),
the distribution of the overlaps or the angle values between these states is
expected to be Gaussian. This is also numerically shown in Fig.\ref{fig:refdistribution}
for 1000 random quantum states (All histograms in the figures are drawn by using 1000 number of states.) generated using random Schmidt basis and coefficients for an eight-qubit star graph state. 

\begin{figure*}[H]
\centering
\subfigure[]{
\label{fig:histschmidt4a}
\includegraphics[width=3in]{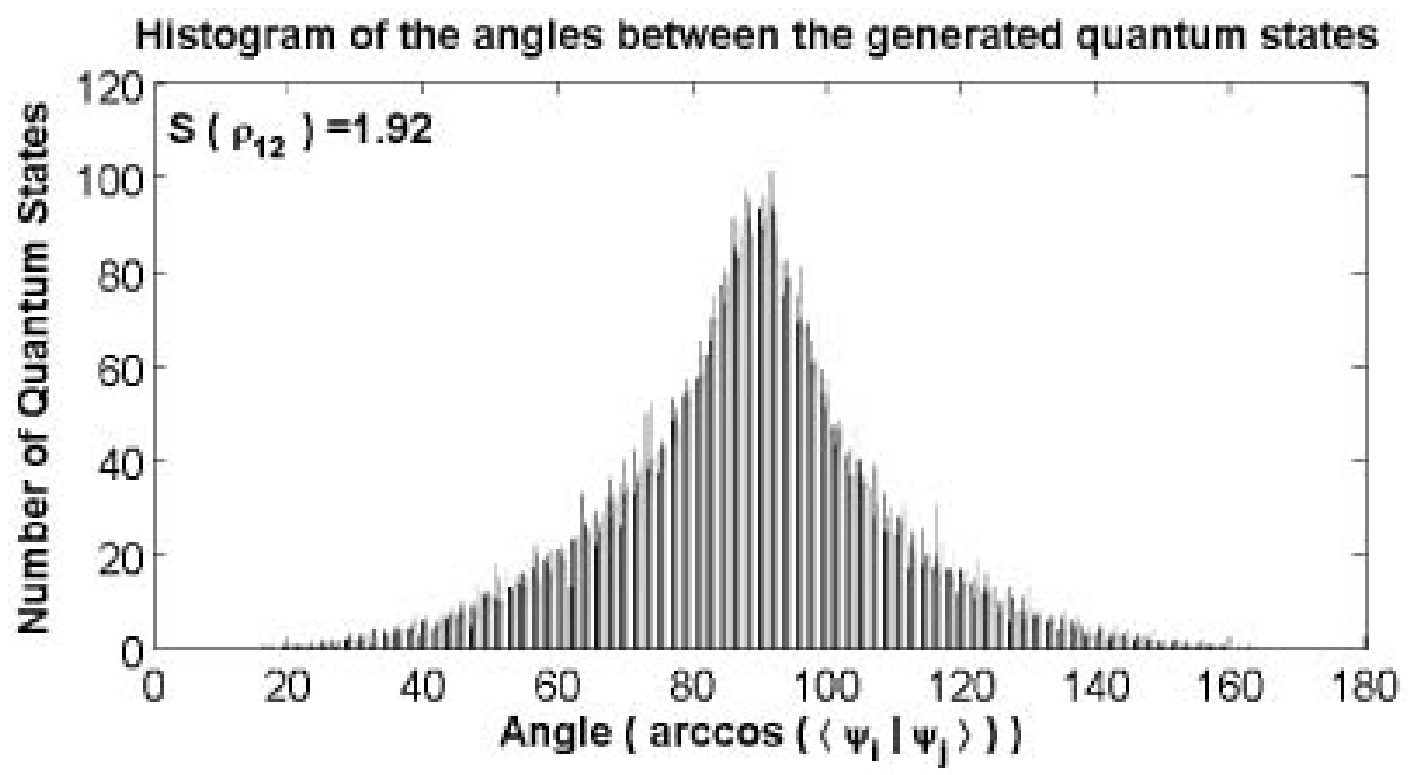}
}\qquad
\subfigure[]{
\label{fig:histschmidt4b}
\includegraphics[width=3in]{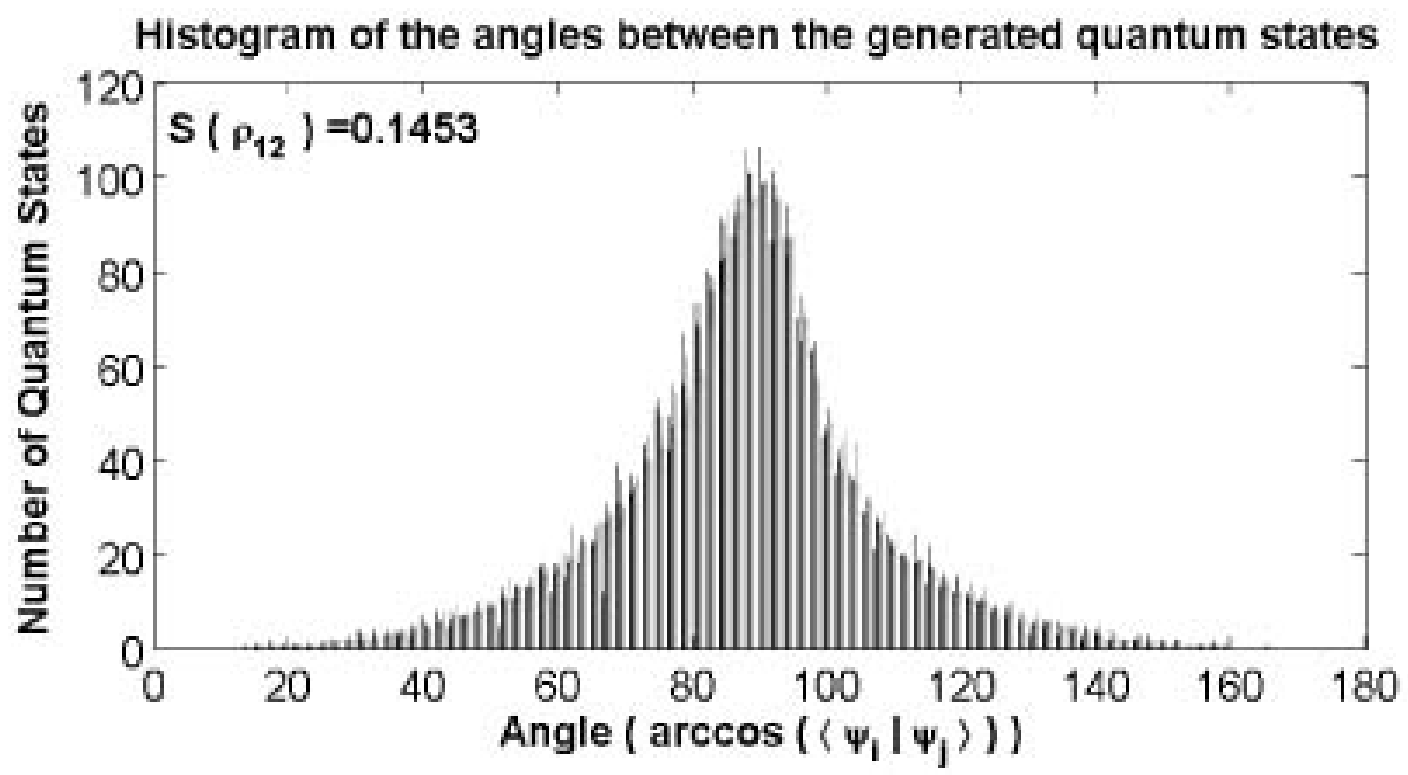}
}
\caption{Histograms of the angles between the generated quantum states for four qubits,
where the Schmidt coefficients are fixed to  certain values and the amount
of the entanglement is high in (a) and low in (b).}
\label{fig:histschmidt4}
\end{figure*}

When we use the same Schmidt coefficients but different random bases to generate random quantum states; if the sizes of the subsystems are greater than one qubit, the distribution of the histogram of the generated quantum states are still Gaussian. This is numerically shown in Fig.\ref{fig:histschmidt4} for a four-qubit system composed of  two-qubit subsystems, $H_{12}$ and $H_{34}$. We draw the distribution of the angles between 1000 random quantum states which has the same Schmidt coefficients but different bases: the comparison of the histograms in Fig.\ref{fig:histschmidt4a} and Fig.\ref{fig:histschmidt4b} shows that the distributions are very similar when the entanglement between subsystem is high and low. Therefore, if the sizes of the subsystems are greater than one qubit, the amount of the bipartite entanglement between subsystems does not affect the distribution, which is Gaussian. 

\begin{figure*}
\centerline{
\resizebox{5in}{!}{    
    \begin{tikzpicture}[thick,stepy=3mm]
    %
    \tikzstyle{operator} = [draw,fill=white,minimum size=1.0em] 
    \tikzstyle{control} = [fill,draw=black!105, shape=circle,minimum size=5pt,inner sep=0pt]	
\tikzstyle{operator2} = [fill=white,draw, shape=circle,minimum size=1em,inner sep=0pt]	
\tikzstyle{operatorbig} = [draw,fill=white,minimum size=4.0em] 
  \tikzstyle{cross} = [fill,draw=black!105, shape=cross out,rotate=45,minimum size=5pt,inner sep=0pt]
\tikzstyle{zerocontrol}= [fill=white,draw=black!105,shape=circle,minimum size=5pt,inner sep=0pt]	
    \tikzstyle{surround} = [fill=blue!10,thick,draw=black,rounded corners=1mm]
    %
    \node[control] at (-0,0) (q1) {};
    \node[control] at (2,0) (q2) {};
    \node[control] at (4,0) (q3) {};
    \node[control] at (6,0) (q4) {};
    \node[control] at (6,-2) (q5) {};    
	\node[control] at (4,-2) (q6) {};
    \node[control] at (2,-2) (q7) {};  
    \node[control] at (0,-2) (q8) {};
\node[] (edge1) at (q4){} edge[-](q1);
\node[] (edge1) at (q8){} edge[-](q5);
\node[] (edge1) at (q5){} edge[-](q4);
\node (c) at (-0.2,-0.2){};
\node at ($(q1)+(c)$ (equiv) {$1$};
\node at ($(q2)+(c)$ (equiv) {$2$};
\node at ($(q3)+(c)$ (equiv) {$3$};
\node at ($(q4)+(c)$ (equiv) {$4$};
\node at ($(q5)+(c)$ (equiv) {$5$};
\node at ($(q6)+(c)$ (equiv) {$6$};
\node at ($(q7)+(c)$ (equiv) {$7$};
\node at ($(q8)+(c)$ (equiv) {$8$};
\node (c) at (1,0.2){};
\node at ($(q1)+(c)$ (equiv) {$.9986$};
\node at ($(q2)+(c)$ (equiv) {$.8362$};
\node at ($(q3)+(c)$ (equiv) {$.2069$};
\node at ($1/2*(q4)+1/2*(q5)+1/2*(c)$ (equiv) {$.5852$};
\node at ($(q6)+(c)$ (equiv) {$.0073$};
\node at ($(q7)+(c)$ (equiv) {$.3010$};
\node at ($(q8)+(c)$ (equiv) {$.7419$};
    \end{tikzpicture}
    \includegraphics[width=3in]{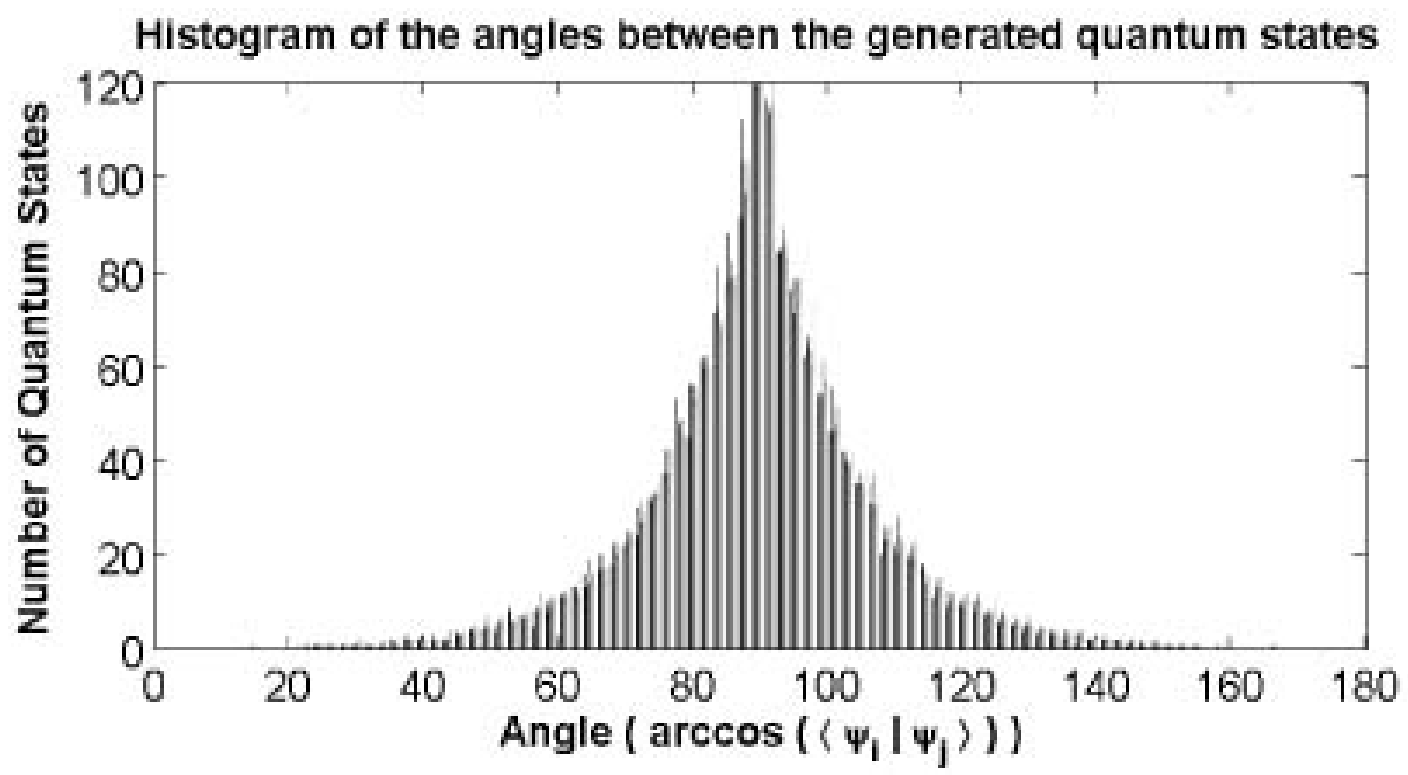}
  }}
\caption{Histogram of the angles between the generated random quantum states for eight qubits where the Schmidt coefficients are fixed  and the corresponding bipartite entanglements are given as weights of the edges on the graph. }
\label{fig:histlineargraph}
\end{figure*}

\begin{figure*}
\centerline{
\resizebox{5in}{!}{
    \begin{tikzpicture}[thick,stepy=3mm]
    %
    \tikzstyle{operator} = [draw,fill=white,minimum size=1.0em] 
    \tikzstyle{control} = [fill,draw=black!105, shape=circle,minimum size=5pt,inner sep=0pt]	
\tikzstyle{operator2} = [fill=white,draw, shape=circle,minimum size=1em,inner sep=0pt]	
\tikzstyle{operatorbig} = [draw,fill=white,minimum size=4.0em] 
  \tikzstyle{cross} = [fill,draw=black!105, shape=cross out,rotate=45,minimum size=5pt,inner sep=0pt]
\tikzstyle{zerocontrol}= [fill=white,draw=black!105,shape=circle,minimum size=5pt,inner sep=0pt]	
    \tikzstyle{surround} = [fill=blue!10,thick,draw=black,rounded corners=1mm]
    %
    \node[control] at (0,0) (q1) {};
    \node[control] at (  1.5637,    1.2470) (q2) {};
    \node[control] at (    1.9499,   -0.4450) (q3) {};
    \node[control] at (   0.8678,   -1.8019) (q4) {};
    \node[control] at (  -0.8678,   -1.8019) (q5) {};    
	\node[control] at (  -1.9499,   -0.4450) (q6) {};
    \node[control] at (   -1.5637,    1.2470) (q7) {};  
    \node[control] at ( -0.0000,    2.0000) (q8) {};
\node[] (edge1) at (q1){} edge[-](q2);
\node[] (edge1) at (q1){} edge[-](q3);
\node[] (edge1) at (q1){} edge[-](q4);
\node[] (edge1) at (q1){} edge[-](q5);
\node[] (edge1) at (q1){} edge[-](q6);
\node[] (edge1) at (q1){} edge[-](q7);
\node[] (edge1) at (q1){} edge[-](q8);
\node (c) at (-0.2,-0.2){};
\node at ($(q1)+(c)$ (equiv) {$1$};
\node at ($(q2)+(c)$ (equiv) {$2$};
\node at ($(q3)+(c)$ (equiv) {$3$};
\node at ($(q4)+(c)$ (equiv) {$4$};
\node at ($(q5)+(c)$ (equiv) {$5$};
\node at ($(q6)+(c)$ (equiv) {$6$};
\node at ($(q7)+(c)$ (equiv) {$7$};
\node at ($(q8)+(c)$ (equiv) {$8$};
\node (c) at (0.2,-0.3){};
\node at ($1/2*(q1)+1/2*(q2)+1/2*(c)$  (equiv) {$ .9618$};
\node at ($1/2*(q1)+1/2*(q3)+1/2*(c)$  (equiv) {$.9671$};
\node at ($1/2*(q1)+1/2*(q4)+1/2*(c)$  (equiv) {$.1971$};
\node at ($1/2*(q1)+1/2*(q5)+1/2*(c)$ (equiv) {$.4376$};
\node at ($1/2*(q1)+1/2*(q6)+1/2*(c)$  (equiv) {$.6751$};
\node at ($1/2*(q1)+1/2*(q7)+1/2*(c)$  (equiv) {$.3629$};
\node at ($1/2*(q1)+1/2*(q8)+1/2*(c)$  (equiv) {$.0655$};
    \end{tikzpicture}\ \ \ 
    \includegraphics[width=3in]{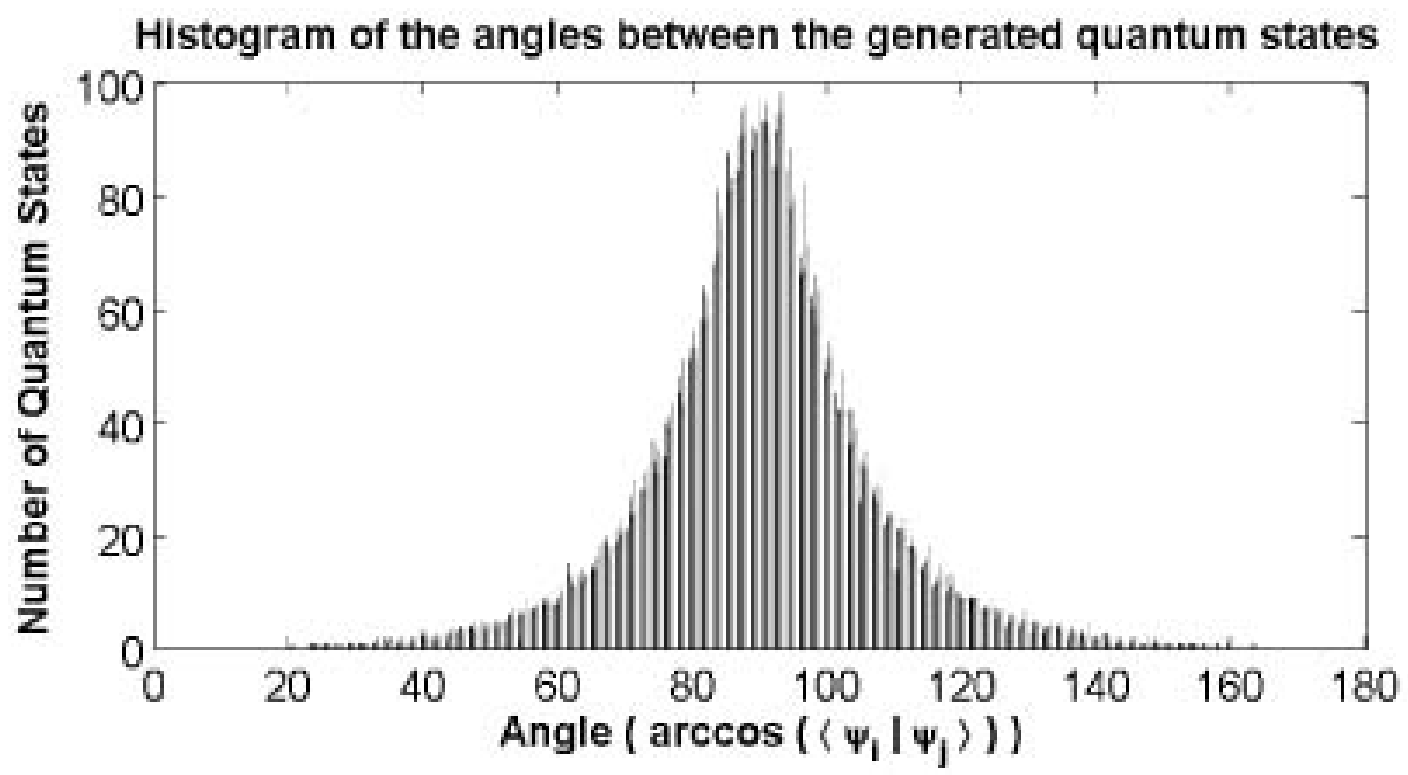}
  }}
\caption{Histogram of the angles between the generated random quantum states for eight qubits where the Schmidt coefficients are fixed  and the corresponding bipartite entanglements are given as weights of the edges on the graph. }
\label{fig:histstargraph}
\end{figure*}

\begin{figure*}[h]
\centering
\subfigure[]{
\label{fig:histschmidt2a}
\includegraphics[width=3in]{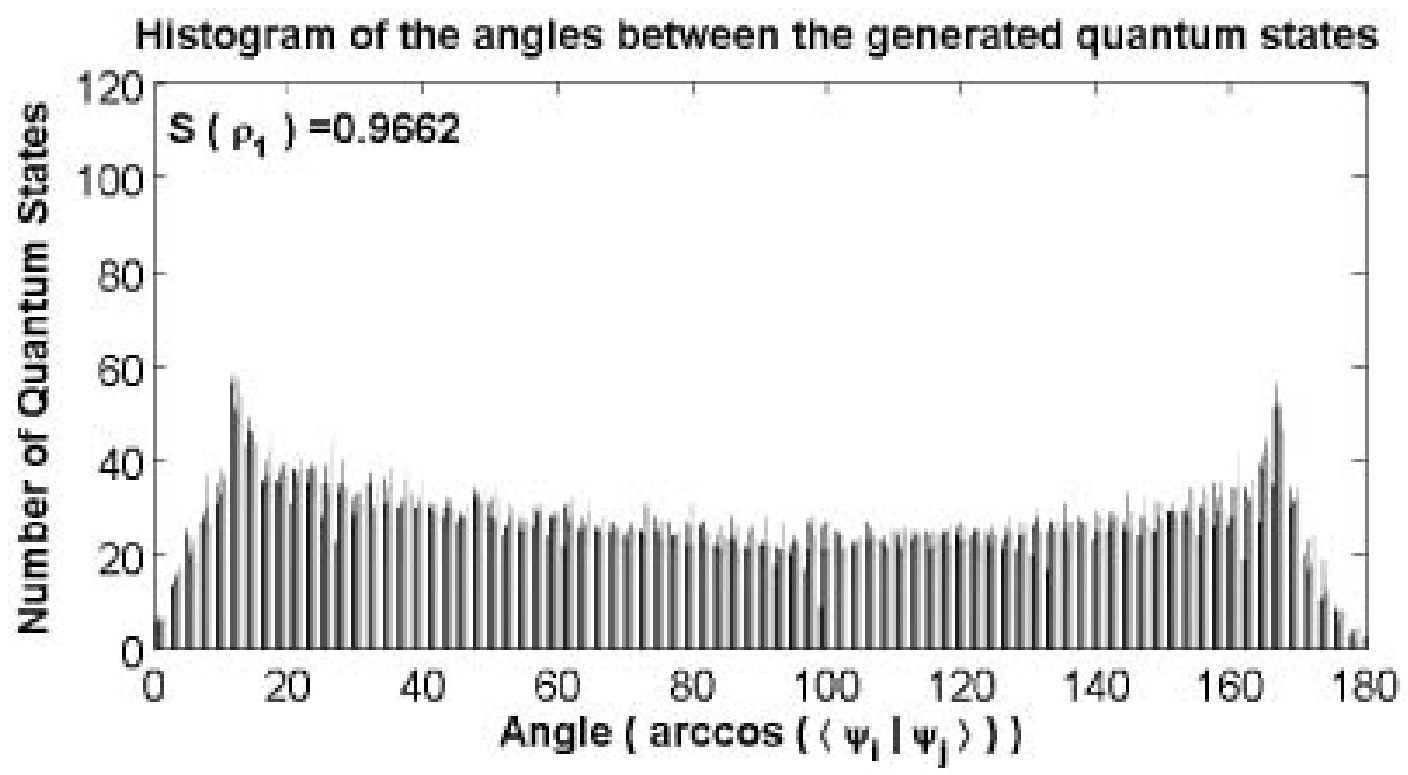}
}\qquad
\subfigure[]{
\label{fig:histschmidt2b}
\includegraphics[width=3in]{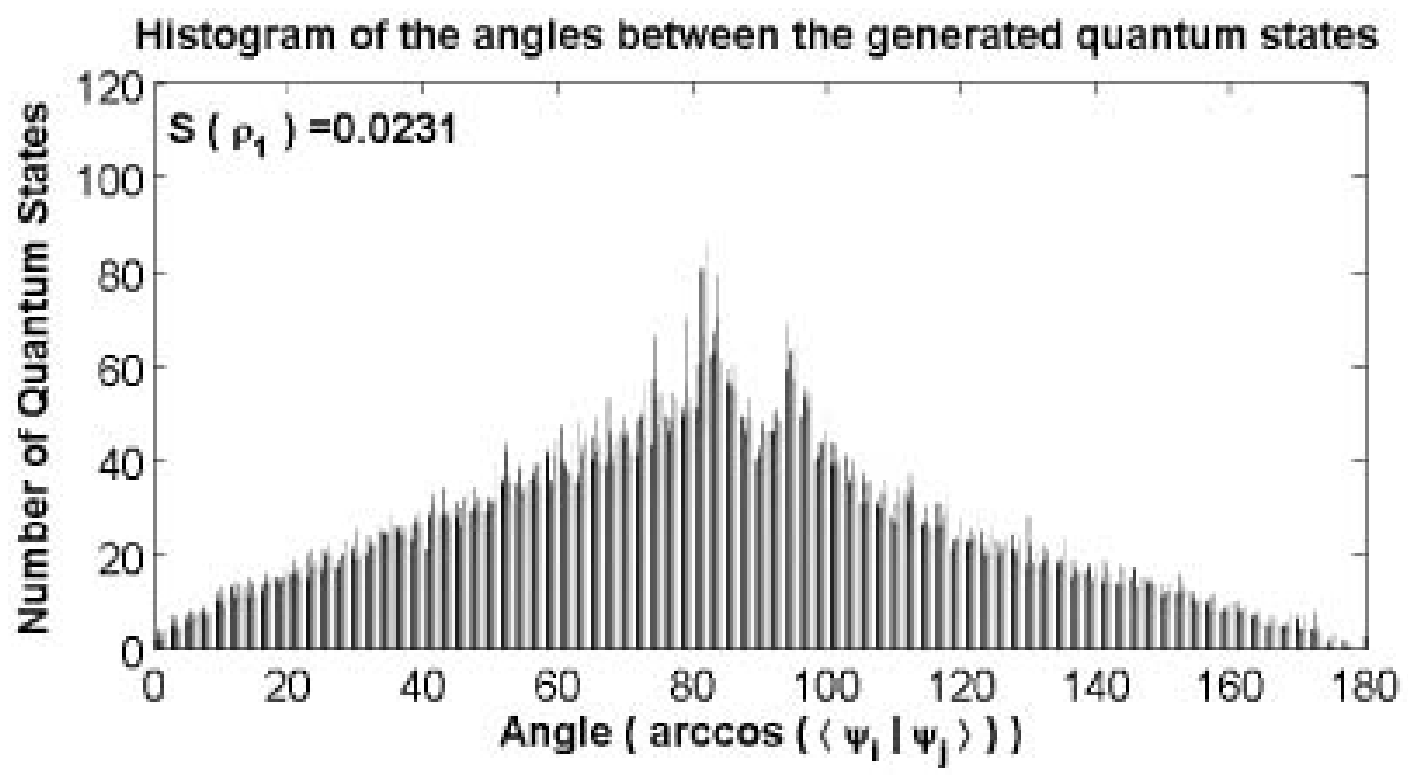}}
\caption{Histograms of the angles between the generated quantum states for two qubits,
where the Schmidt coefficients are fixed to a certain value and the amount
of the entanglement is high in (a) and low in (b).}
\label{fig:histschmidt2}
\end{figure*}

\begin{figure*}[H]
\centering
\subfigure[$\overline{S}=0.0838$]{
\includegraphics[width=2in]{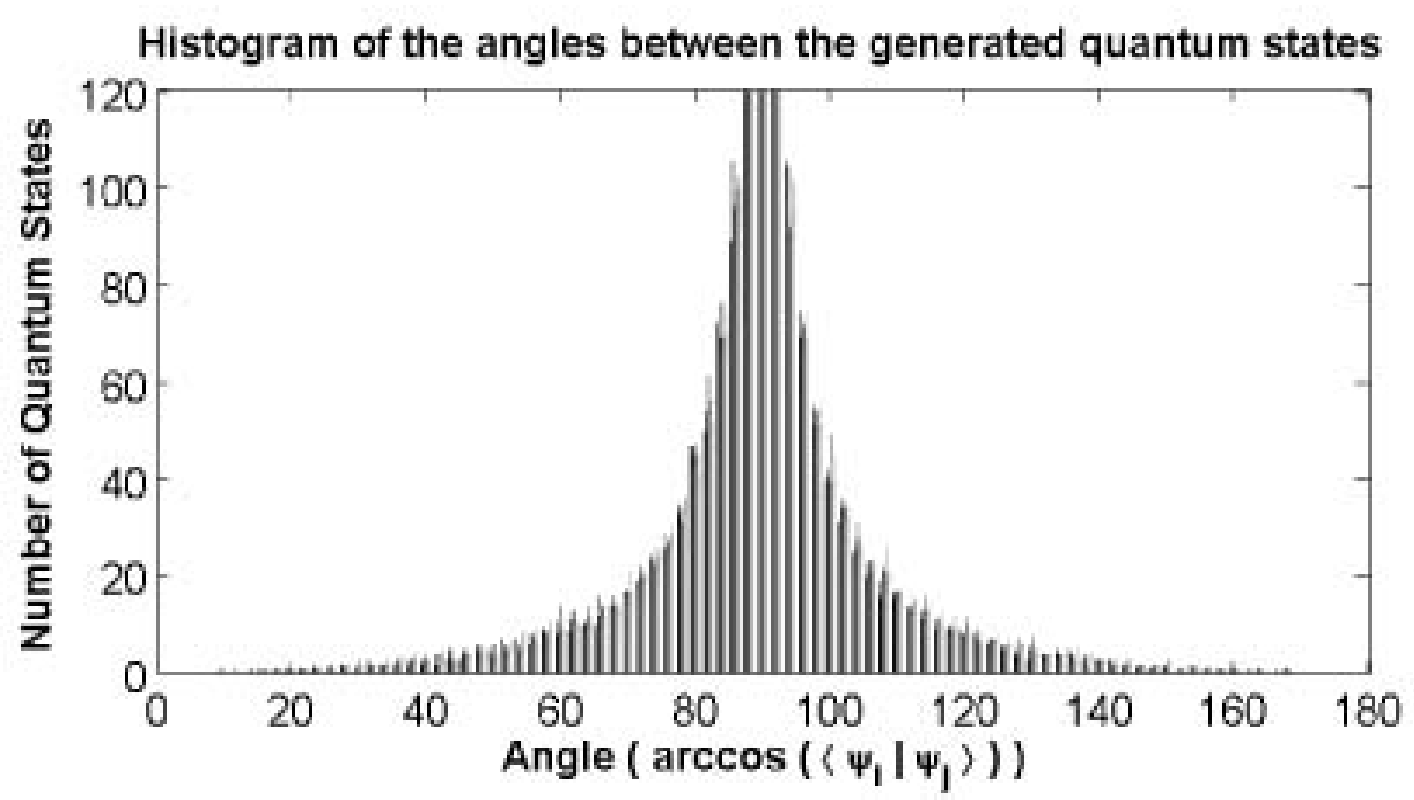}
}\qquad
\subfigure[$\overline{S}=0.1974$]{
\includegraphics[width=2in]{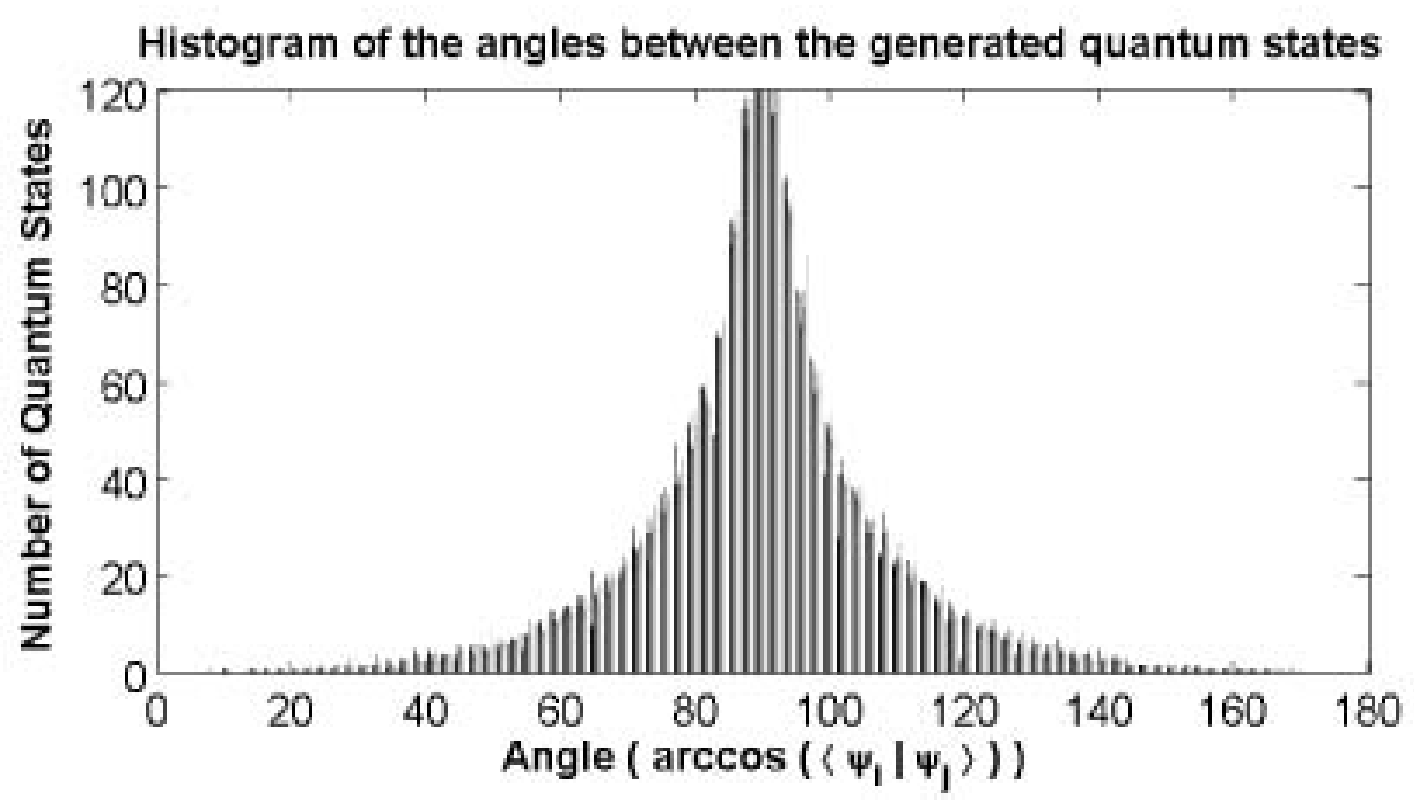}
}\qquad
\subfigure[$\overline{S}=0.3012$]{
\includegraphics[width=2in]{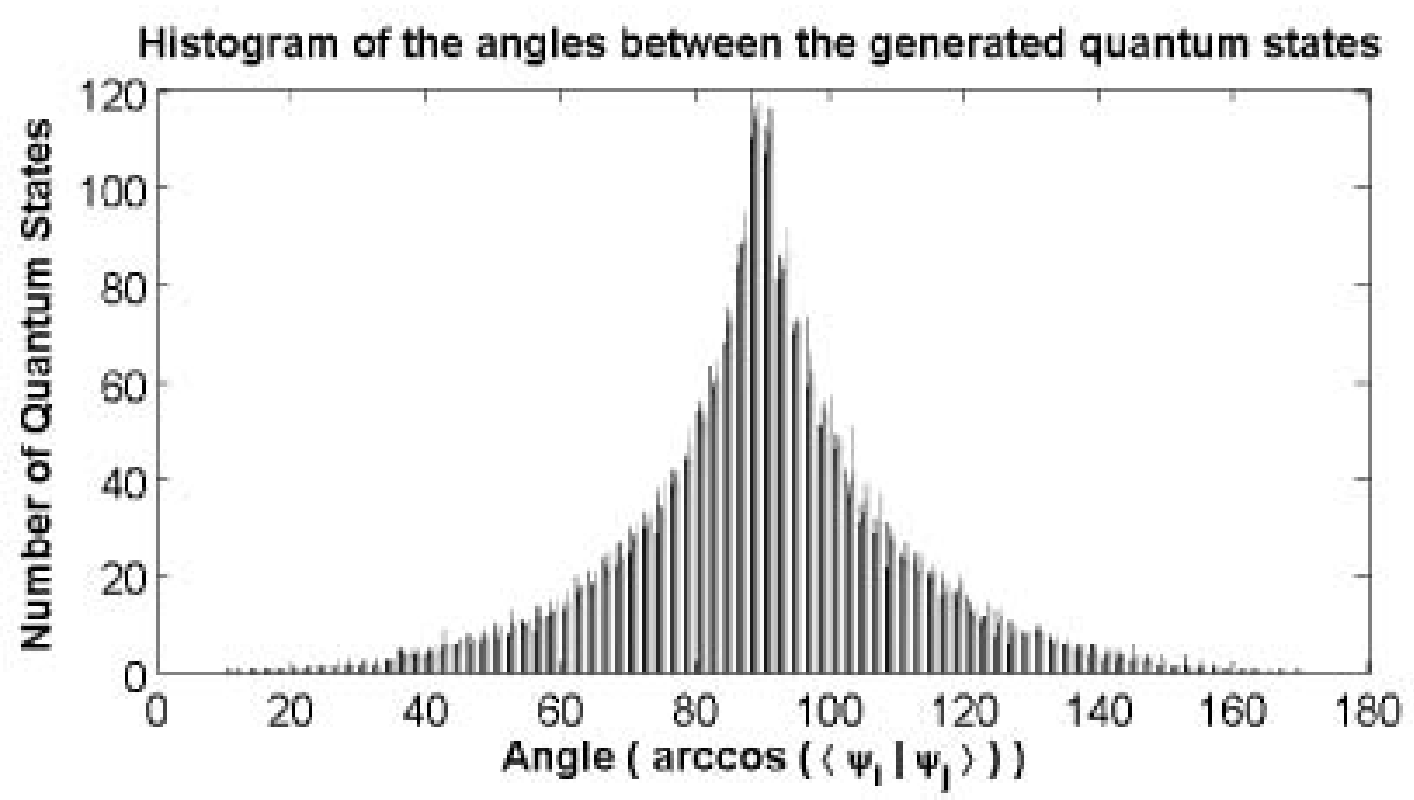}
}\qquad
\subfigure[$\overline{S}=0.3887$]{
\includegraphics[width=2in]{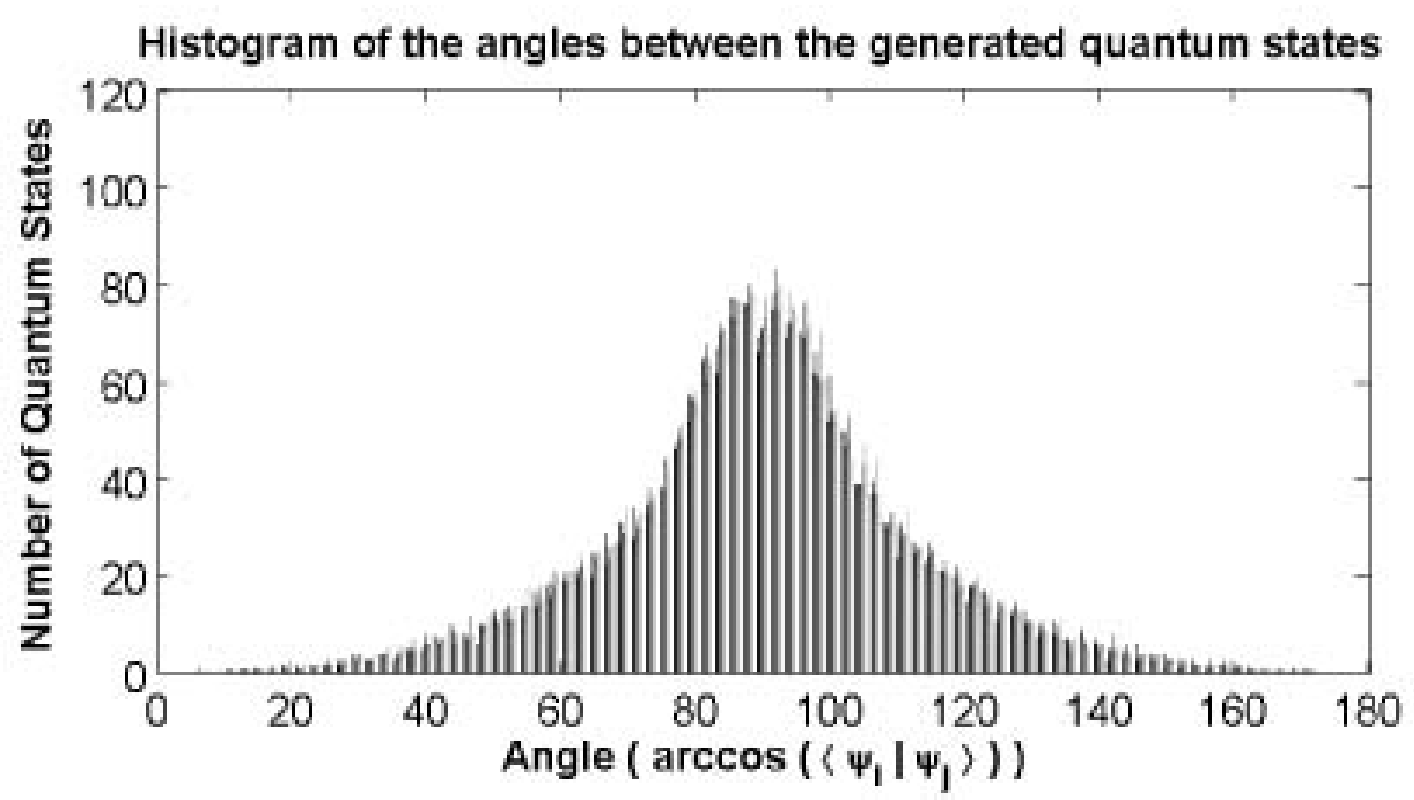}
}\qquad
\subfigure[$\overline{S}=0.4735$]{
\includegraphics[width=2in]{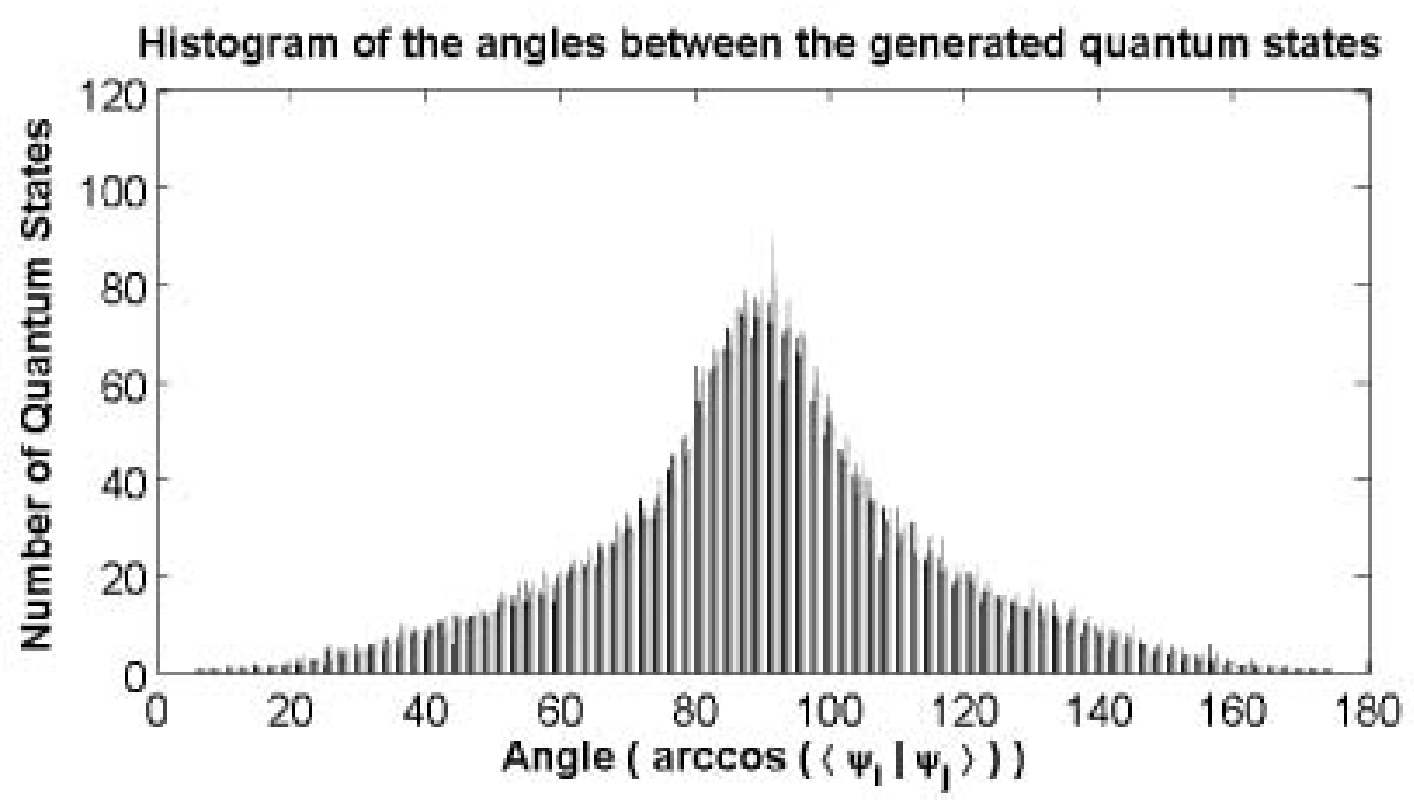}
}\qquad
\subfigure[$\overline{S}=0.5723$]{
\includegraphics[width=2in]{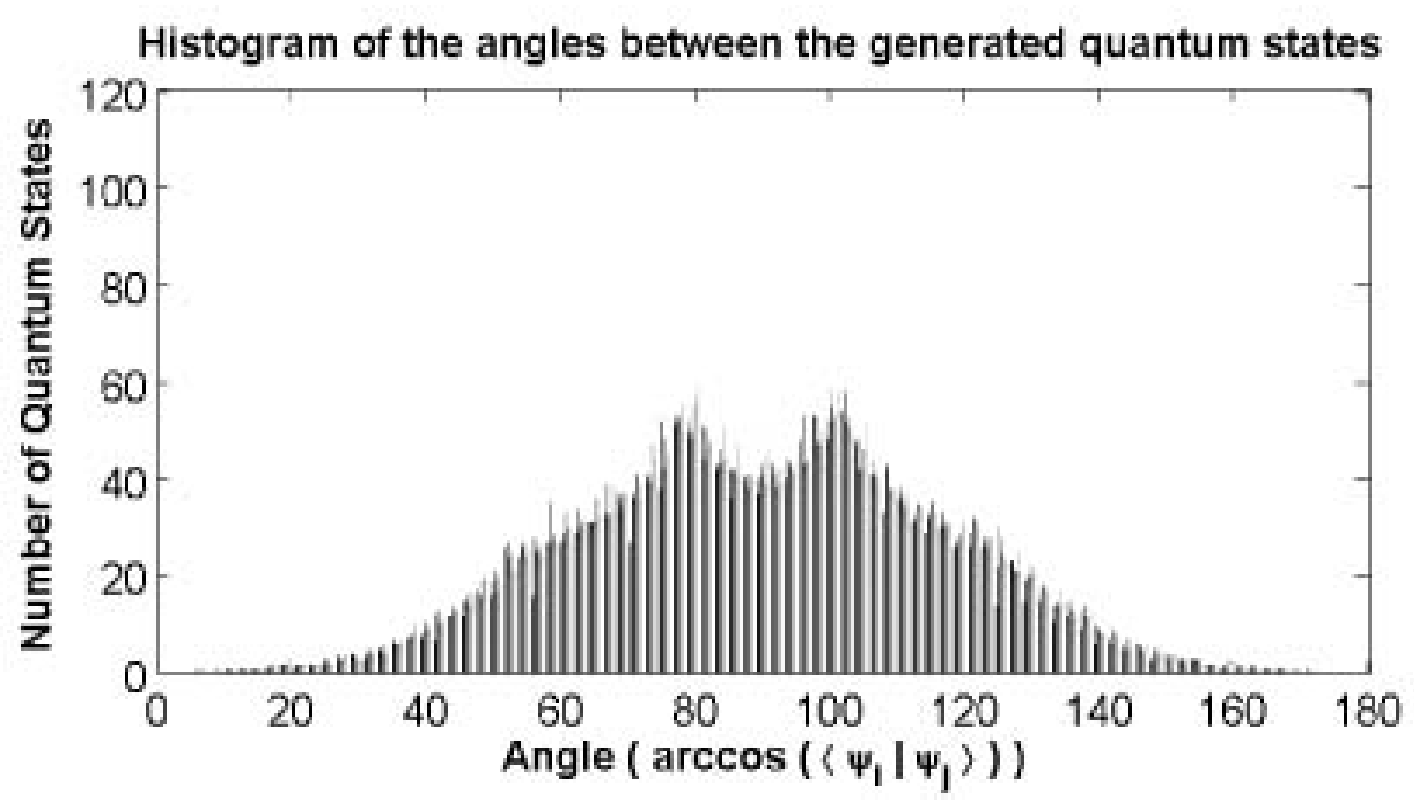}
}\qquad
\subfigure[$\overline{S}=0.6839$]{
\includegraphics[width=2in]{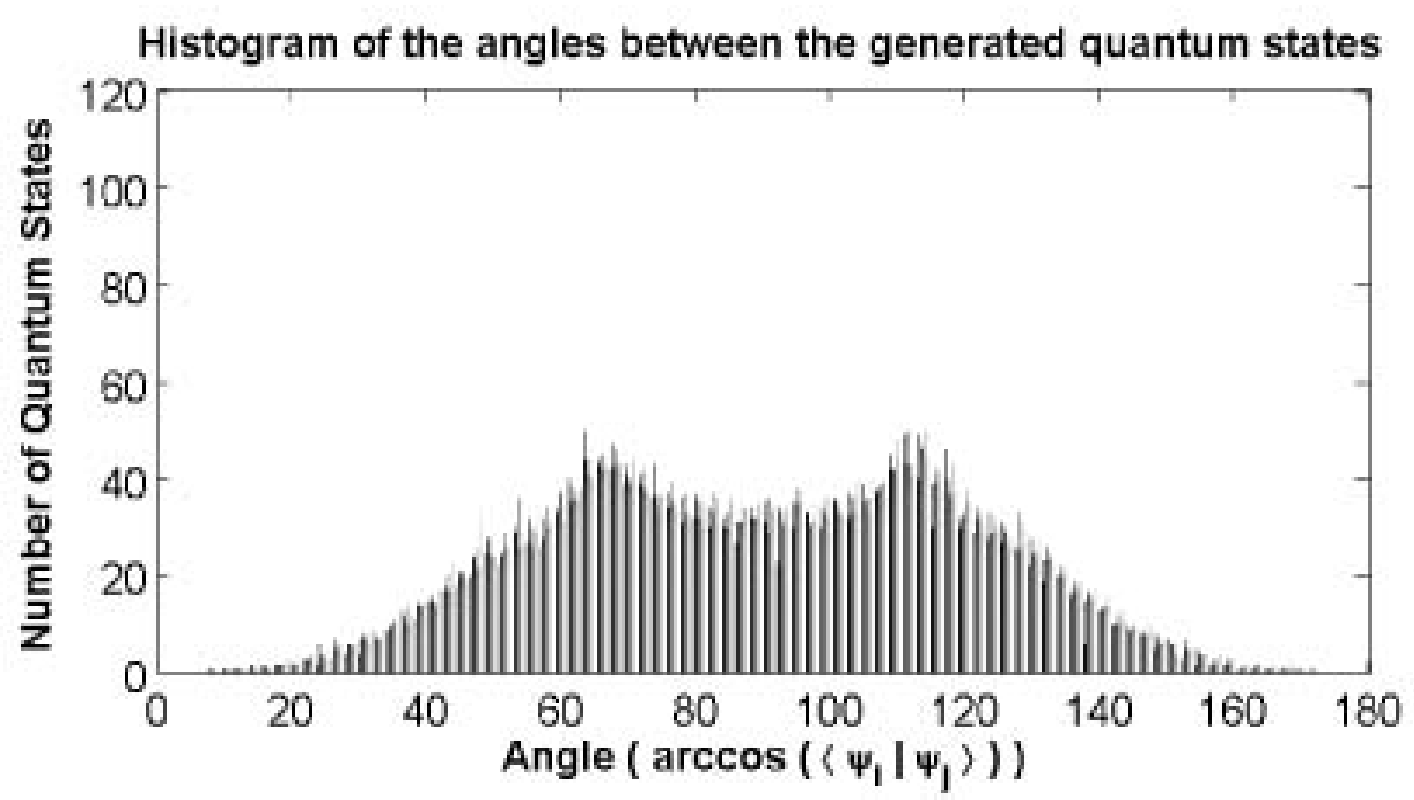}
}\qquad
\subfigure[$\overline{S}=0.7655$]{
\includegraphics[width=2in]{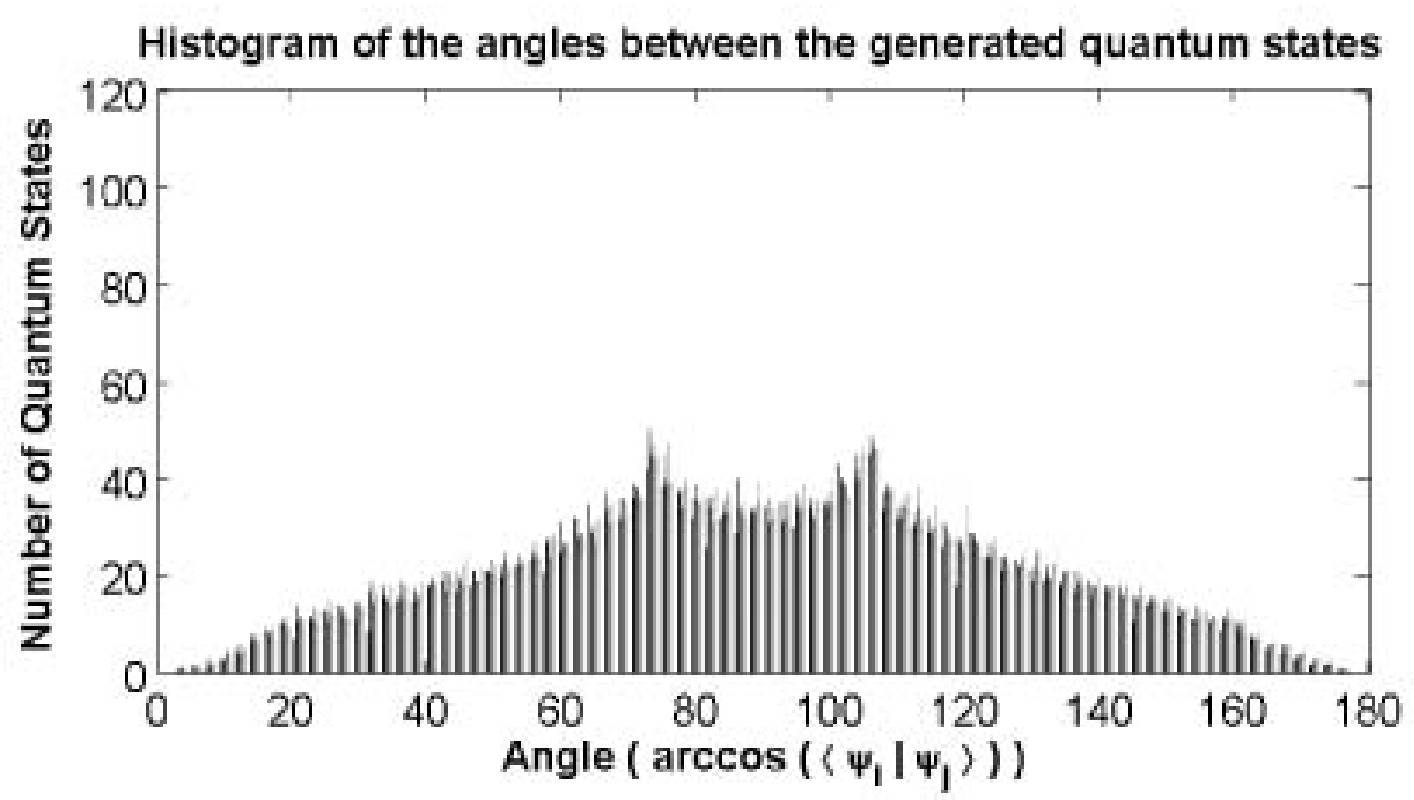}
}\qquad
\subfigure[$\overline{S}=0.8578$]{
\includegraphics[width=2in]{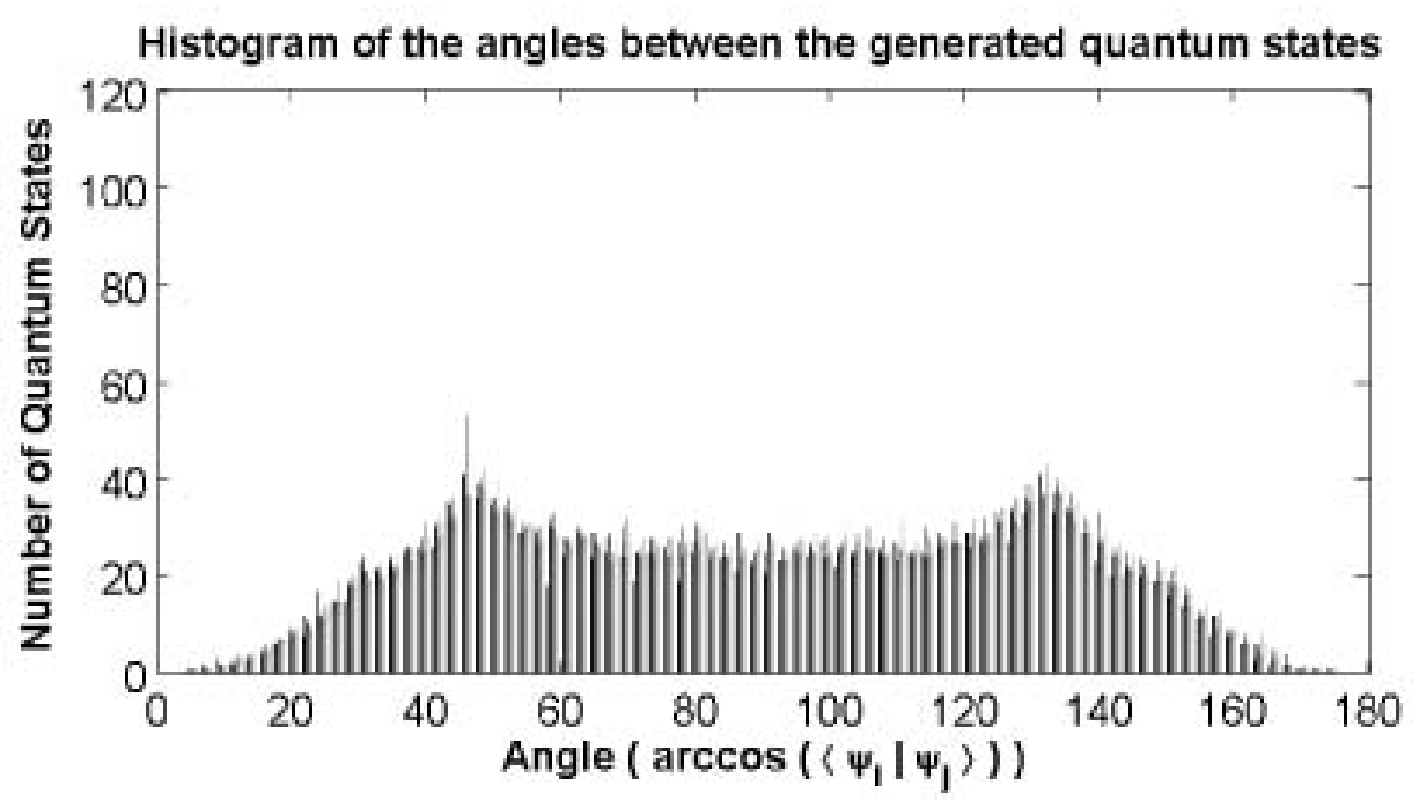}
}\qquad
\subfigure[$\overline{S}=0.9440$]{
\includegraphics[width=2in]{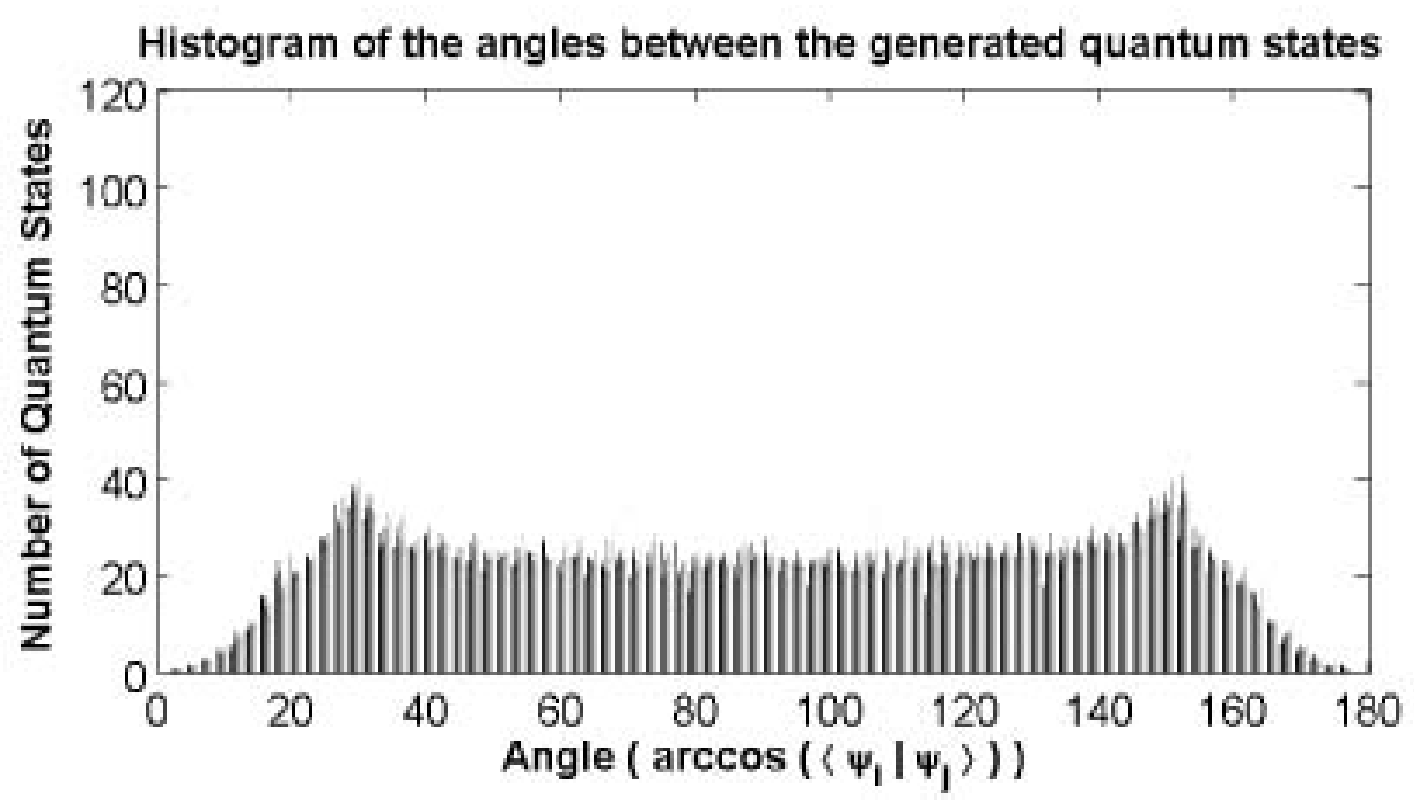}
}
\caption{Histograms of the angles between the generated quantum states for a 5-qubit star graph, where the Schmidt coefficients are fixed to certain values for each figure and the amount of the average bipartite entanglement ($\overline{S}$)  is given under each figure. }
\label{fig:histschmidt5qubits}
\end{figure*}

\begin{figure*}[H]
\centering
\subfigure[3-qubit: $\overline{S}=0.5755$]{
\includegraphics[width=2in]{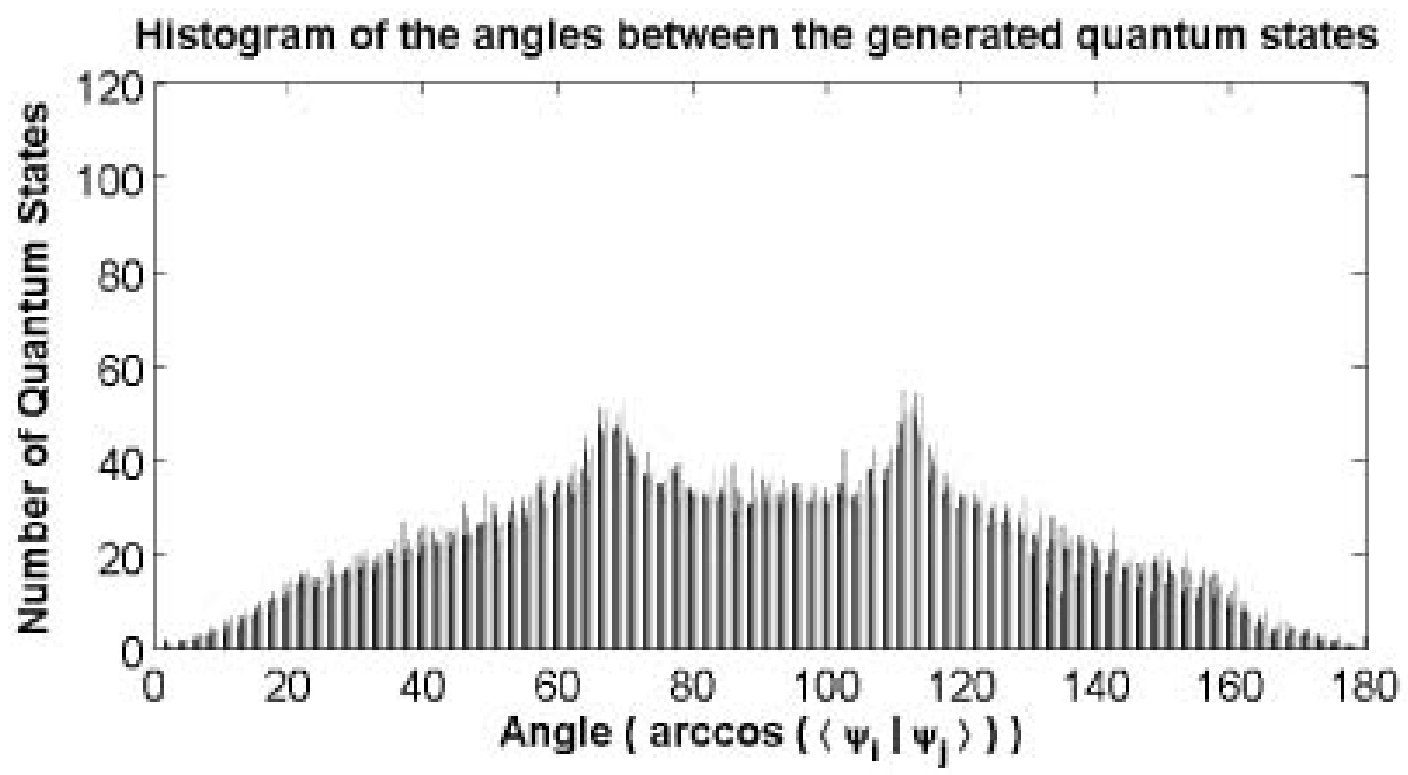}
}
\subfigure[4-qubit: $\overline{S}=0.2691$]{
\includegraphics[width=2in]{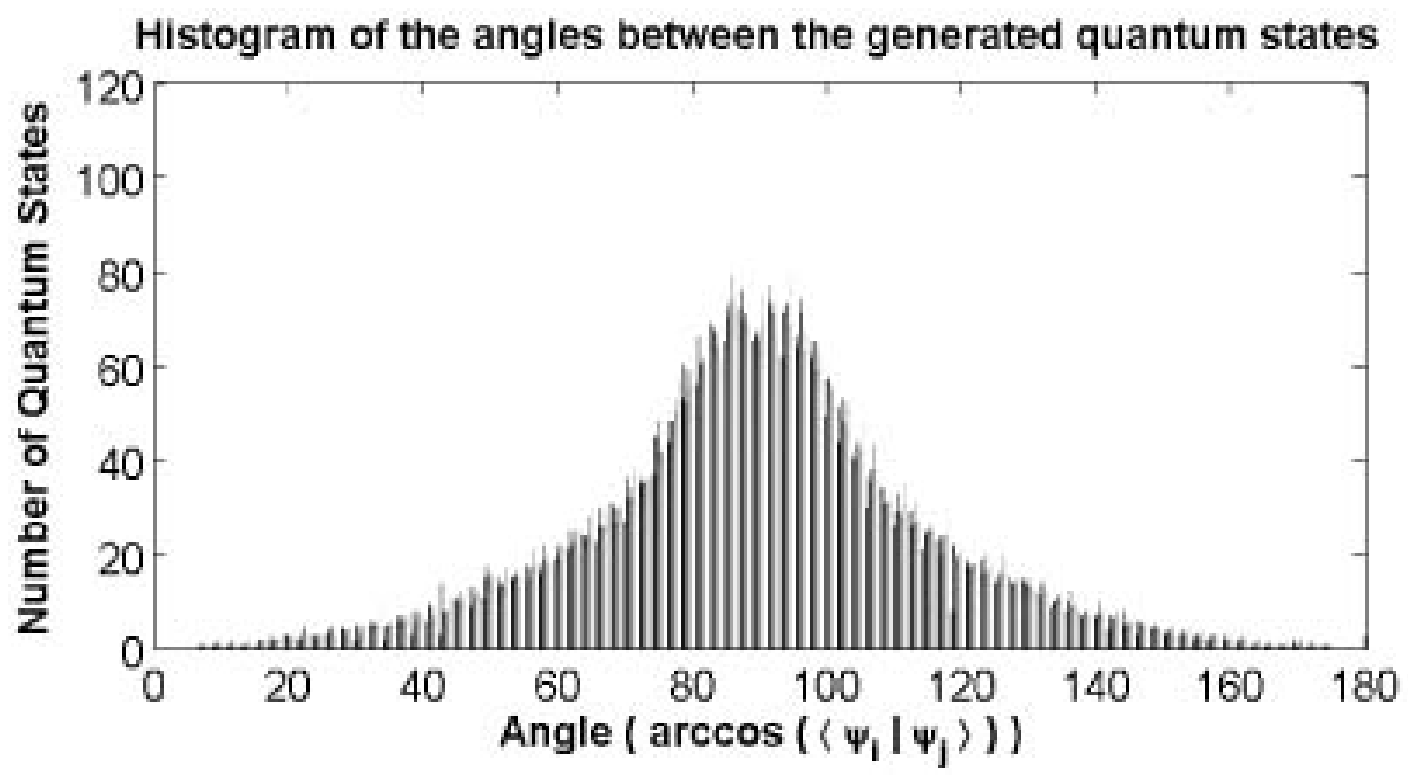}
}\qquad
\subfigure[5-qubit: $\overline{S}=0.6791$]{
\includegraphics[width=2in]{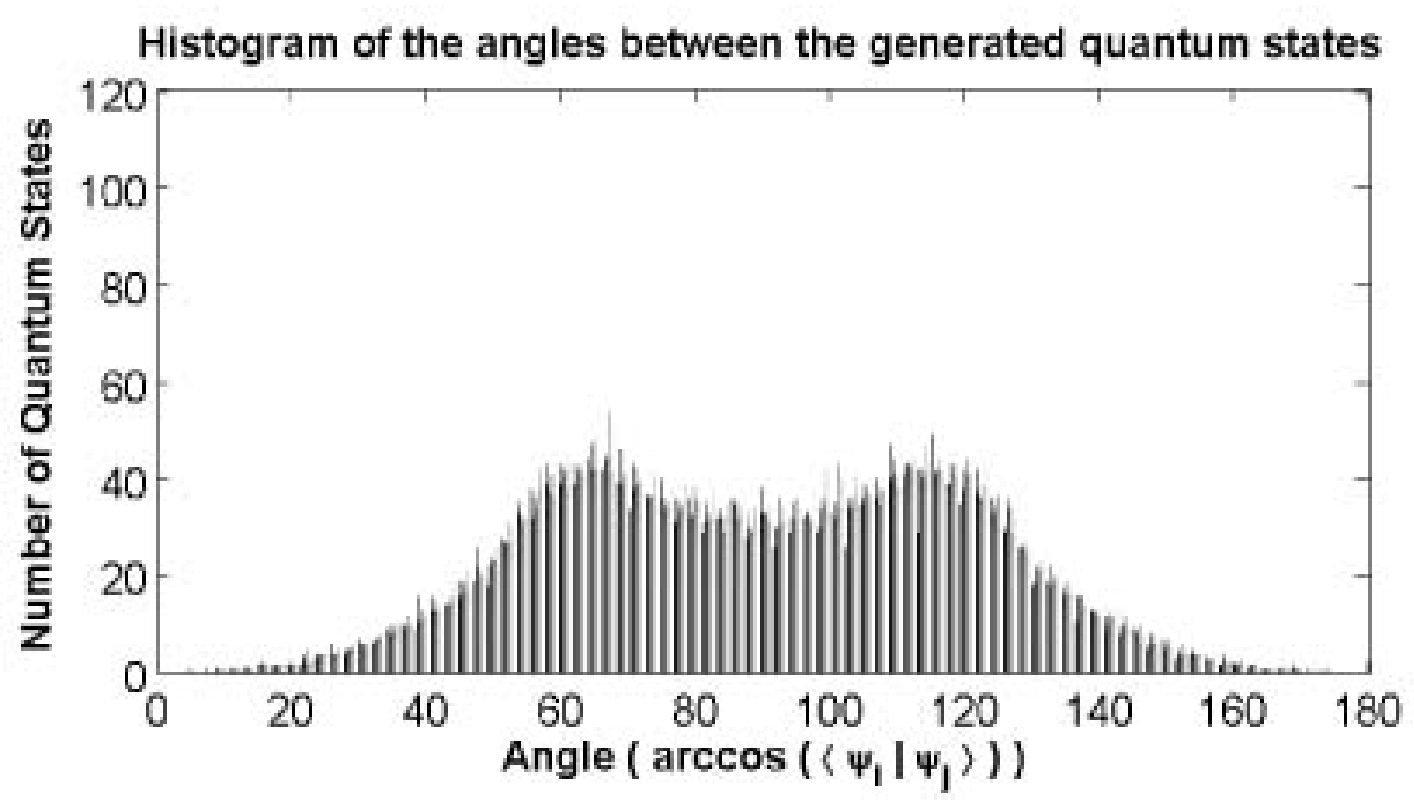}
}\qquad
\subfigure[6-qubit: $\overline{S}=0.4676$]{
\includegraphics[width=2in]{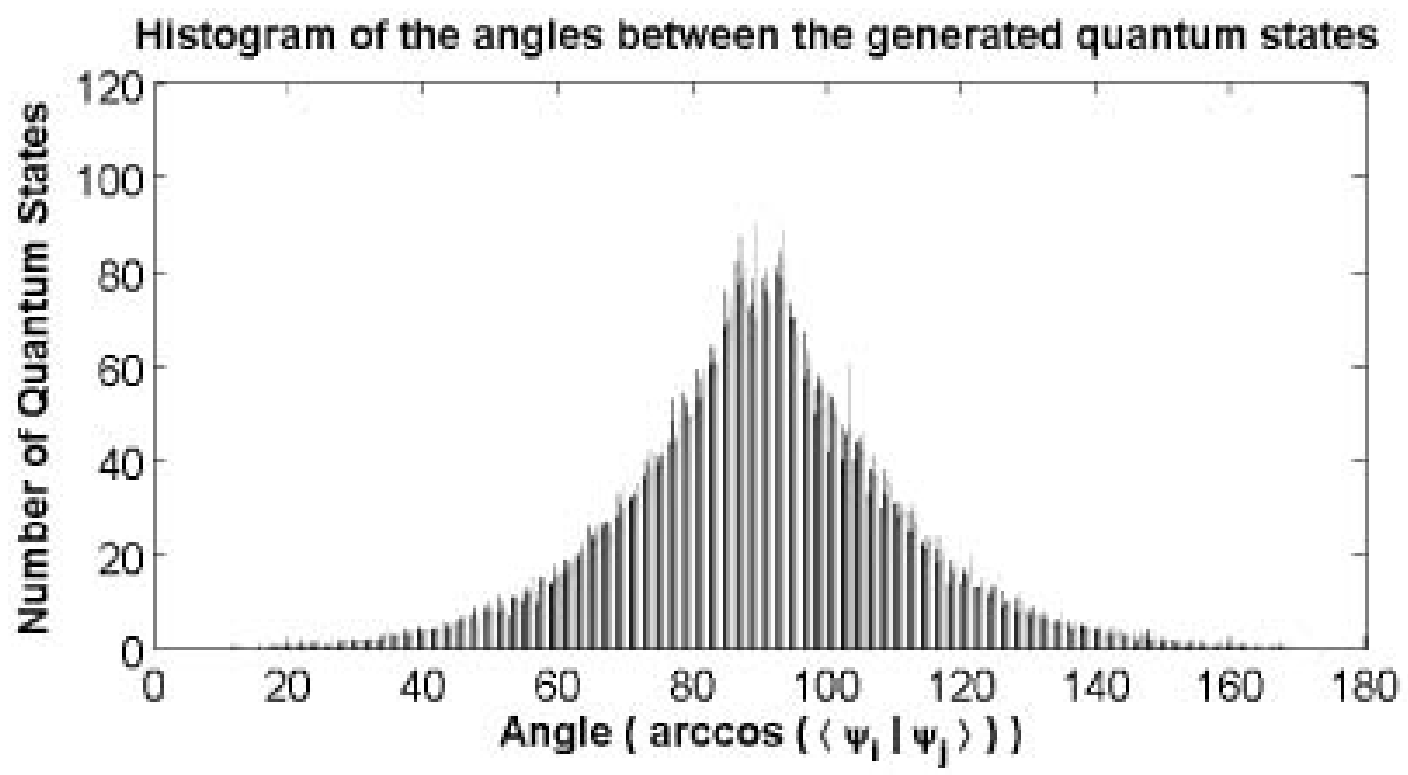}
}\qquad
\subfigure[7-qubit: $\overline{S}=0.8564$]{
\includegraphics[width=2in]{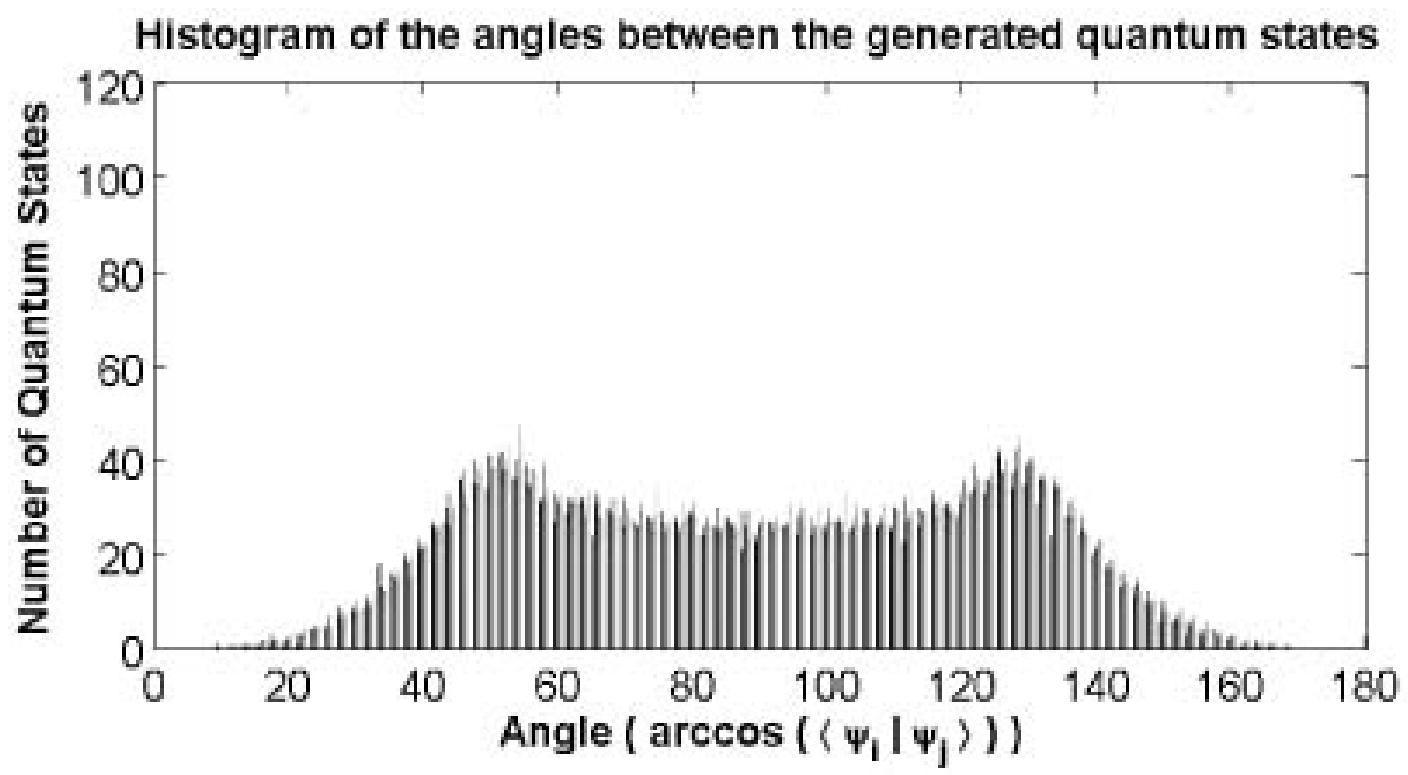}
}\qquad
\subfigure[8-qubit: $\overline{S}=0.6462$]{
\includegraphics[width=2in]{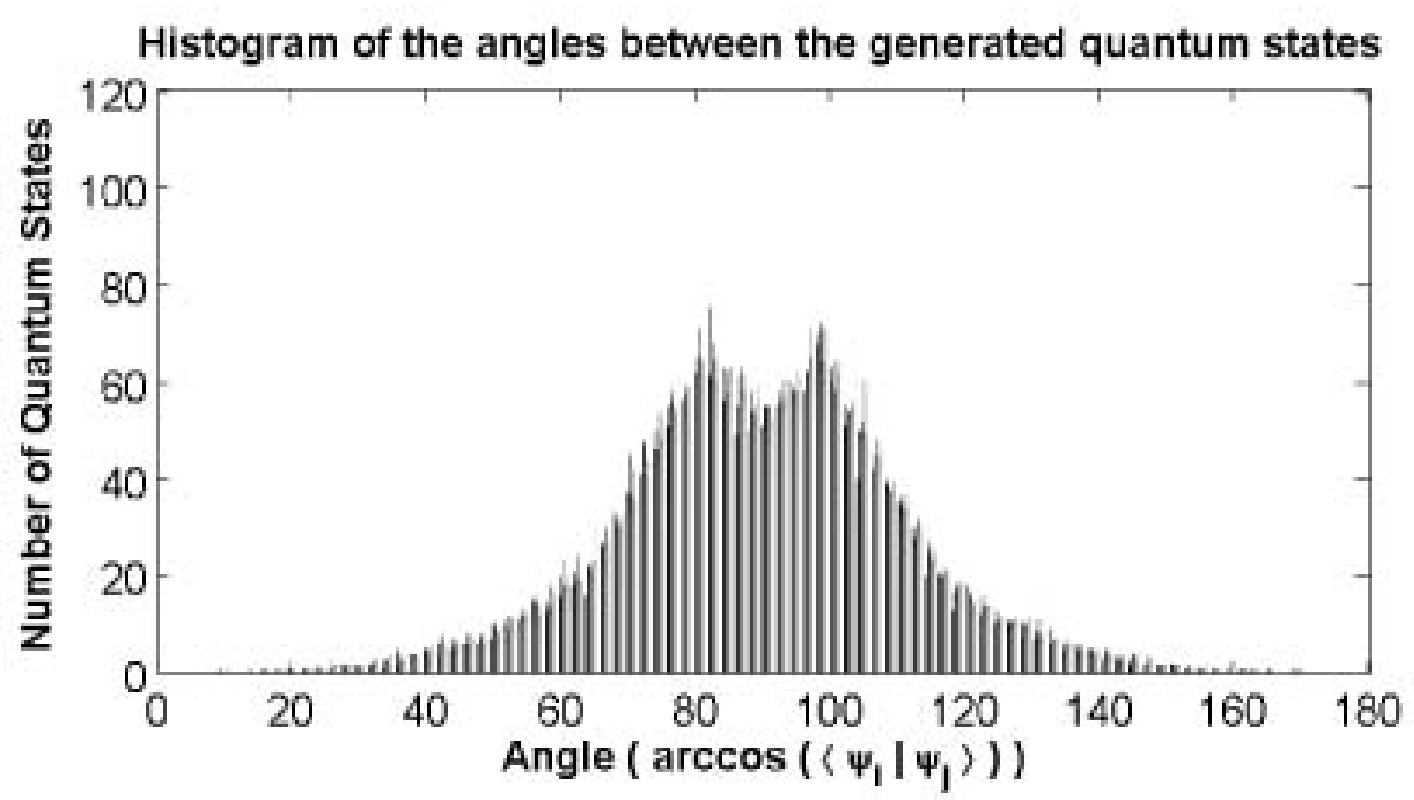}
}\qquad
\subfigure[9-qubit: $\overline{S}=0.5337$]{
\includegraphics[width=2in]{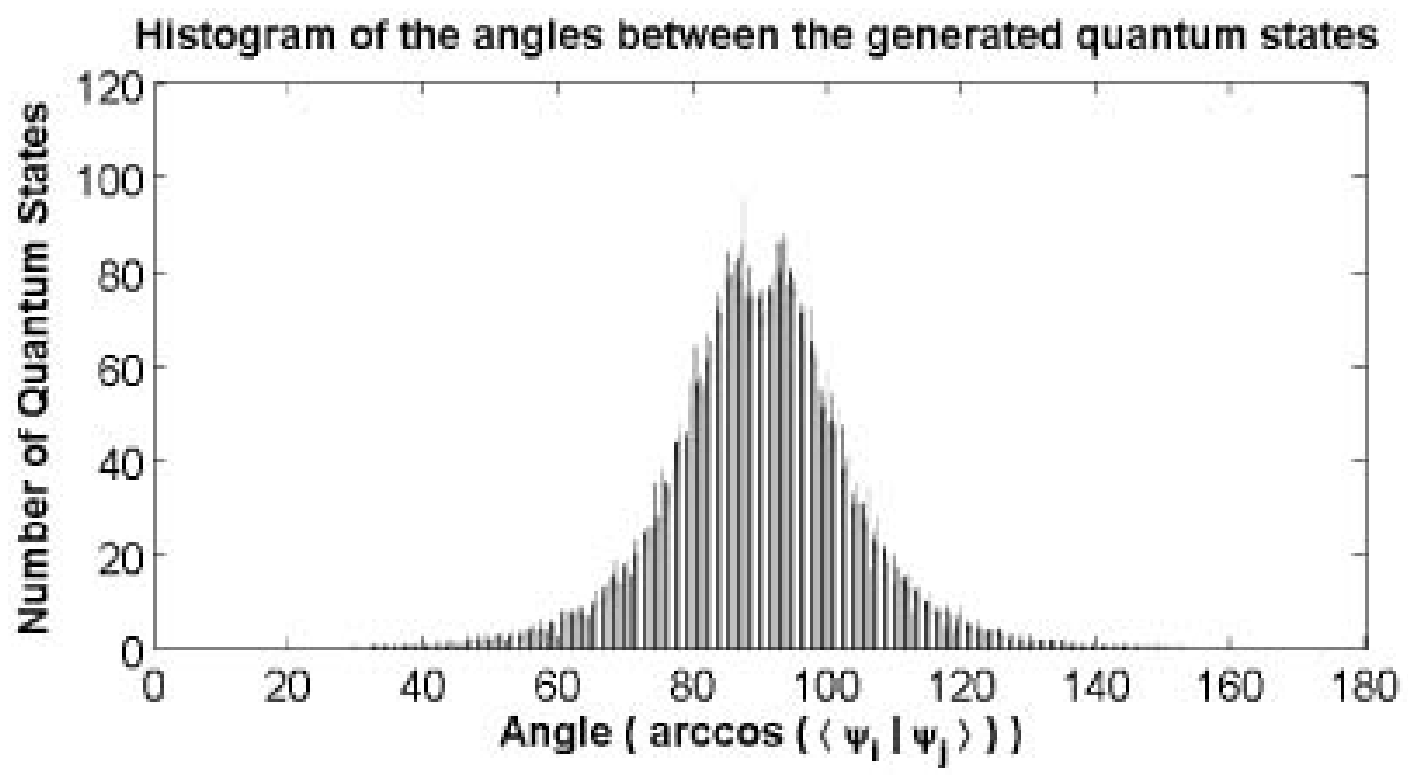}
}\qquad
\subfigure[10-qubit: $\overline{S}=0.6766$]{
\includegraphics[width=2in]{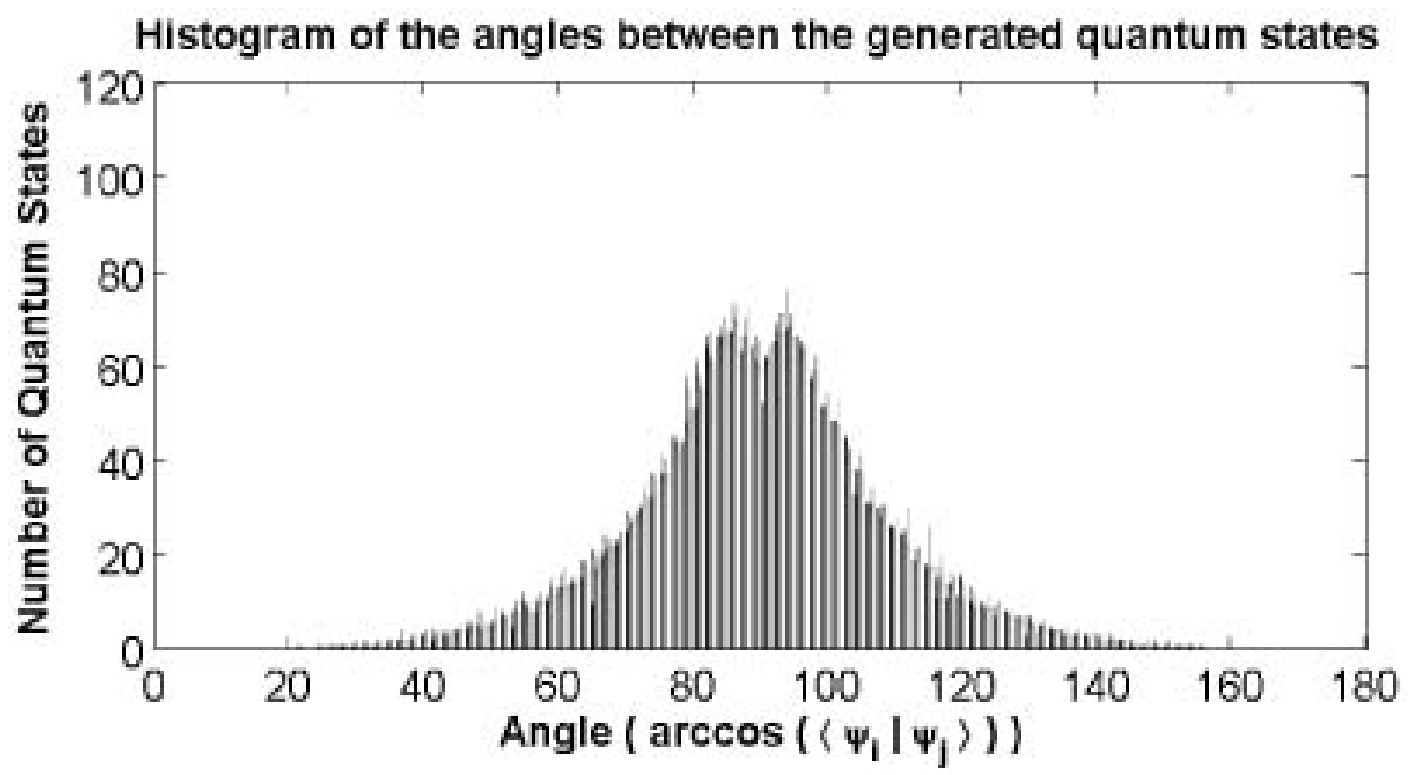}
}
\caption{Histograms of the angles between the generated quantum states for different number of qubits, where
the Schmidt coefficients are fixed to certain values and the amount of the average bipartite entanglement ($\overline{S}$) is given under each figure. }
\label{fig:histschmidtdifferentqubits}
\end{figure*}

However, if the size of one of the subsystems is one qubit and the amount of the entanglement is fixed, examples are shown in Fig.\ref{fig:histlineargraph} for an eight-qubit linear graph and in Fig.\ref{fig:histstargraph} for an eight-qubit star graph, then the distribution of the angles between the generated quantum states are affected by the amount of the bipartite entanglement between these subsystems. As an example in Fig.\ref{fig:histschmidt2} for a two-qubit system,  two different random set of Schmidt coefficients are used to generate two group of 1000 quantum random states (the states in the same group have the same Schmidt coefficients): the entanglement is high for the first group and low for the second group. 
While the histogram of the first group shown in Fig.\ref{fig:histschmidt2a} looks  more uniform-like, the histogram for the second group shown in Fig.\ref{fig:histschmidt2b} is more Gaussian-like. 
This indicates that the distribution changes by the amount of the entanglement and is Gaussian when the entanglement is low; however, becomes more uniform-like when the amount of the entanglement is increased.  
This appears more clearly in Fig.\ref{fig:histschmidt5qubits}, where the average bipartite entanglement ($\overline{S}$) between subsystems of a five-qubit star graph changes for each different group of 1000 random states. As shown in Fig.\ref{fig:histschmidt5qubits}, the distribution becomes more uniform-like when the average entanglement ($\overline{S}$) is increased.  Please also note that the reason for using five qubits is to perform the computer simulations faster while keeping the system size large enough. In addition, we  show the histograms of star graphs of different number of qubits with different average bipartite entanglements in Fig.\ref{fig:histschmidtdifferentqubits}, which  also supports Fig.\ref{fig:histschmidt5qubits} and the above argument.

\section{Conclusion}

In this paper, we map the Schmidt decomposition for a general quantum state into
quantum circuits, which can be used to generate random quantum states.
We show that in random state generation, the entanglement amount between
subsystems can be controlled by using quantum gates implementing the desired
Schmidt coefficients. We also show that one can combine the Schmidt circuits
sequentially to generate an equivalent quantum state for an acyclic weighted
graph state in which vertices and edges correspond to subsystems and the
bipartite entanglement between subsystems, respectively, and also the
amount of entanglement is given by the weights of the edges. 
  
Our method can be used in different applications and protocols relying on entanglement.
In the simulation of quantum systems, one can use the method to create an instance
of the desired system. In addition, decoherence effects the quality of the
entanglement and generally cause errors in computations.  A similar idea
can be used in quantum error correction to correct an imperfect bipartite
entanglement, and so a quantum channel.




\begin{thebibliography}{100}
\providecommand{\url}[1]{{#1}}
\providecommand{\urlprefix}{URL }
\expandafter\ifx\csname urlstyle\endcsname\relax
  \providecommand{\doi}[1]{DOI \discretionary{}{}{}#1}\else
  \providecommand{\doi}{DOI \discretionary{}{}{}\begingroup
  \urlstyle{rm}\Url}\fi

\bibitem{Edelman2005}
A.~Edelman, N.R. Rao, Acta Numerica \textbf{14}, 233 (2005).
\newblock \doi{10.1017/S0962492904000236}

\bibitem{Tulino2004random}
A.M. Tulino, S.~Verd{\'u}, \emph{Random matrix theory and wireless
  communications}, vol.~1 (Now Publishers Inc, 2004)

\bibitem{Beenakker1997}
C.W.J. Beenakker, Rev. Mod. Phys. \textbf{69}, 731 (1997).
\newblock \doi{10.1103/RevModPhys.69.731}

\bibitem{Wiesner1983}
S.~Wiesner, SIGACT News \textbf{15}(1), 78 (1983)

\bibitem{Wootters1990random}
W.K. Wootters, Foundations of Physics \textbf{20}(11), 1365 (1990)

\bibitem{Sakurai1985modern}
J.J. Sakurai, \emph{Modern quantum mechanics} (Reading, MA: Addison
  Wesley,|edited by Tuan, San Fu, 1985)

\bibitem{Hans2001}
H.J. Briegel, R.~Raussendorf, Phys. Rev. Lett. \textbf{86}, 910 (2001).
\newblock \doi{10.1103/PhysRevLett.86.910}

\bibitem{Dur2003}
W.~D\"ur, H.~Aschauer, H.J. Briegel, Phys. Rev. Lett. \textbf{91}, 107903
  (2003).
\newblock \doi{10.1103/PhysRevLett.91.107903}

\bibitem{Nest2004}
M.~Van~den Nest, J.~Dehaene, B.~De~Moor, Phys. Rev. A \textbf{69}, 022316
  (2004).
\newblock \doi{10.1103/PhysRevA.69.022316}

\bibitem{Otfried2005}
O.~G\"uhne, G.~T\'oth, P.~Hyllus, H.J. Briegel, Phys. Rev. Lett. \textbf{95},
  120405 (2005).
\newblock \doi{10.1103/PhysRevLett.95.120405}

\bibitem{Scarani2005}
V.~Scarani, A.~Acin, E.~Schenck, M.~Aspelmeyer, Phys. Rev. A \textbf{71},
  042325 (2005).
\newblock \doi{10.1103/PhysRevA.71.042325}

\bibitem{Hein2004}
M.~Hein, J.~Eisert, H.J. Briegel, Phys. Rev. A \textbf{69}, 062311 (2004).
\newblock \doi{10.1103/PhysRevA.69.062311}

\bibitem{Schlingemann2001}
D.~Schlingemann, R.F. Werner, Phys. Rev. A \textbf{65}, 012308 (2001).
\newblock \doi{10.1103/PhysRevA.65.012308}

\bibitem{Raussendorf2001}
R.~Raussendorf, H.J. Briegel, Phys. Rev. Lett. \textbf{86}, 5188 (2001).
\newblock \doi{10.1103/PhysRevLett.86.5188}

\bibitem{HeinReview}
M.~Hein, W.~Dür, J.~Eisert, R.~Raussendorf, M.~Van~den Nest, H.J. Briegel, in
  the Proceedings of the International School of Physics "Enrico Fermi" on
  "Quantum Computers, Algorithms and Chaos", Varenna, Italy, July \textbf{162},
  115 (2006).
\newblock \doi{10.3254/978-1-61499-018-5-115}

\bibitem{lu2007experimental}
C.Y. Lu, X.Q. Zhou, O.~G{\"u}hne, W.B. Gao, J.~Zhang, Z.S. Yuan, A.~Goebel,
  T.~Yang, J.W. Pan, Nature Physics \textbf{3}(2), 91 (2007)

\bibitem{Hartmann2005}
W.~D\"ur, L.~Hartmann, M.~Hein, M.~Lewenstein, H.J. Briegel, Phys. Rev. Lett.
  \textbf{94}, 097203 (2005).
\newblock \doi{10.1103/PhysRevLett.94.097203}

\bibitem{Hartmann2007}
L.~Hartmann, J.~Calsamiglia, W.~Dür, H.J. Briegel, Journal of Physics B:
  Atomic, Molecular and Optical Physics \textbf{40}(9), S1 (2007)

\bibitem{Anders2006}
S.~Anders, M.B. Plenio, W.~D\"ur, F.~Verstraete, H.J. Briegel, Phys. Rev. Lett.
  \textbf{97}, 107206 (2006).
\newblock \doi{10.1103/PhysRevLett.97.107206}

\bibitem{Anders2007}
S.~Anders, H.J. Briegel, W.~Dür, New Journal of Physics \textbf{9}(10), 361
  (2007)

\bibitem{Campbell2007}
E.T. Campbell, J.~Fitzsimons, S.C. Benjamin, P.~Kok, Phys. Rev. A \textbf{75},
  042303 (2007).
\newblock \doi{10.1103/PhysRevA.75.042303}

\bibitem{White1999}
A.G. White, D.F. James, P.H. Eberhard, P.G. Kwiat, Physical review letters
  \textbf{83}(16), 3103 (1999)

\bibitem{Briegel2001}
H.J. Briegel, R.~Raussendorf, Phys. Rev. Lett. \textbf{86}, 910 (2001).
\newblock \doi{10.1103/PhysRevLett.86.910}

\bibitem{Whaley2010}
M.~Sarovar, A.~Ishizaki, G.R. Fleming, K.B. Whaley, Nature Physics
  \textbf{6}(6), 462 (2010)

\bibitem{Fassioli2010}
F.~Fassioli, A.~Olaya-Castro, New Journal of Physics \textbf{12}(8), 085006
  (2010)

\bibitem{Daskin2012}
A.~Daskin, A.~Grama, G.~Kollias, S.~Kais, The Journal of Chemical Physics
  \textbf{137}(23), 234112 (2012).
\newblock \doi{http://dx.doi.org/10.1063/1.4772185}

\bibitem{Facchi2008}
P.~Facchi, U.~Marzolino, G.~Parisi, S.~Pascazio, A.~Scardicchio, Phys. Rev.
  Lett. \textbf{101}, 050502 (2008).
\newblock \doi{10.1103/PhysRevLett.101.050502}.
\newblock
  \urlprefix\url{http://link.aps.org/doi/10.1103/PhysRevLett.101.050502}

\bibitem{Pasquale2010}
A.~De~Pasquale, P.~Facchi, G.~Parisi, S.~Pascazio, A.~Scardicchio, Phys. Rev. A
  \textbf{81}, 052324 (2010).
\newblock \doi{10.1103/PhysRevA.81.052324}.
\newblock \urlprefix\url{http://link.aps.org/doi/10.1103/PhysRevA.81.052324}

\bibitem{Facchi2013}
P.~Facchi, G.~Florio, G.~Parisi, S.~Pascazio, K.~Yuasa, Phys. Rev. A
  \textbf{87}, 052324 (2013).
\newblock \doi{10.1103/PhysRevA.87.052324}.
\newblock \urlprefix\url{http://link.aps.org/doi/10.1103/PhysRevA.87.052324}

\bibitem{Stewart1980}
G.~Stewart, SIAM Journal on Numerical Analysis \textbf{17}(3), 403 (1980)

\bibitem{Anderson1987}
T.W. Anderson, I.~Olkin, L.~Underhill, SIAM Journal on Scientific and
  Statistical Computing \textbf{8}(4), 625 (1987)

\end{thebibliography}
\end{document}